\documentclass[useAMS,usenatbib]{mn2e}
\usepackage[utf8]{inputenc}
\usepackage{epsfig,graphicx,float}
\usepackage{amsmath}
\usepackage{amssymb}
\usepackage{dsfont}
\usepackage{threeparttable} 
\usepackage{booktabs}
\usepackage{xcolor}

\usepackage{epstopdf}
\usepackage{times}

\newcommand{\chan}{\textit{Chandra}}
\newcommand{\swift}{\textit{Swift}}

\newcommand{\xmm}{\textit{XMM-Newton}}

\newcommand{\keck}{\textit{Keck}}

\newcommand{\Msun}{\mathrm{M}_{\odot}}
\newcommand{\lum}{\mathrm{erg~s}^{-1}}
\newcommand{\flux}{\mathrm{erg~cm}^{-2}~\mathrm{s}^{-1}}

\newcommand{\sgra}{Sgr~A$^{*}$}
\def \mnras {MNRAS}
\def \apj {ApJ}
\def \apjs {ApJS}
\def \apjl {ApJL}
\def \aap {A\&A}
\def \nat {Nature}

\def \pasj {PASJ}

\title[Swift study of X-ray flaring in \sgra]{A Swift study of long-term changes in the X-ray flaring properties of Sagittarius A*}
\author[A. Andrés et al.]
{A. Andrés$^{1,2,3}$, J. van den Eijnden$^{4,1}$\thanks{e-mail: jakob.vandeneijnden@physics.ox.ac.uk}, N. Degenaar$^{1}$, P.A. Evans$^{5}$, K. Chatterjee$^{1,6,7}$, \newauthor M. Reynolds$^{8}$, J.M.~Miller$^{8}$,
J.~Kennea$^{9}$, R. Wijnands$^{1}$, S. Markoff$^{1}$, D. Altamirano$^{10}$, \newauthor C.O. Heinke$^{11}$, A. Bahramian$^{12}$, 
G. Ponti$^{13,14}$, D. Haggard$^{15}$ \\
$^1$Anton Pannekoek Institute for Astronomy, University of Amsterdam, Science Park 904, 1098 XH, Amsterdam, the Netherlands\\
$^2$ Instituto de Astronomía, Universidad Nacional Autonoma de México, Av. Universidad 3000, C.U., Coyoacán, 04510 Ciudad de México, CDMX \\
$^3$Facultad de Ciencias Naturales y Matemática, Universidad de El Salvador, Final 25 Avenida Norte, San Salvador, El Salvador\\
$^4$Department of Physics, Astrophysics, University of Oxford, Denys Wilkinson Building, Keble Road, Oxford OX1 3RH, UK\\
$^5$University of Leicester, School of Physics and Astronomy, University Road, Leicester LE1 7RH, UK\\
$^6$ Black Hole Initiative at Harvard University, 20 Garden Street, Cambridge, MA 02138, USA \\
$^7$ Harvard-Smithsonian Center for Astrophysics, 60 Garden Street, Cambridge, MA 02138, USA \\
$^8$Department of Astronomy, University of Michigan, 1085 South University Avenue, Ann Arbor, MI  48109, USA\\
$^9$Department of Astronomy and Astrophysics, 525 Davey Lab, Pennsylvania State University, University Park, PA 16802, USA\\
$^{10}$Department of Physics and Astronomy, University of Southampton, Southampton, Hampshire, SO171BJ, UK\\ 
$^{11}$Department of Physics, University of Alberta, 4-183 CCIS, Edmonton, AB T6G 2E1, Canada\\
$^{12}$International Centre for Radio Astronomy Research – Curtin University, GPO Box U1987, Perth, WA 6845, Australia\\
$^{13}$Osservatorio Astronomico di Brera, Via E. Bianchi 46, I-23807 Merate (LC), Italy\\
$^{14}$Max-Planck-Institut f\"ur Extraterrestrische Physik, Giessenbachstrasse, D-85748, Garching, Germany\\
$^{15}$Department of Physics, McGill University, 3600 rue University, Montreal, QC, H3A 2T8, Canada
}

\begin{document}

\date{DRAFT VERSION}

\pagerange{\pageref{firstpage}--\pageref{lastpage}} \pubyear{0000}

\maketitle

\label{firstpage}

\begin{abstract}
The radiative counterpart of the supermassive black hole at the Galactic Centre, Sagittarius A*, displays flaring emission in the X-ray band atop a steady, quiescent level. Flares are also observed in the near-infrared band. The physical process producing the flares is not fully understood and it is unclear if the flaring rate varies, although some recent works suggest it has reached unprecedented variability in recent years. Using over a decade of regular X-ray monitoring of Neil Gehrels  \swift\ Observatory, we studied the variations in count rate of \sgra\ on time scales of years. We decomposed the X-ray emission into quiescent and flaring emission, modelled as a constant and power law process, respectively. We found that the complete, multi-year dataset cannot be described by a stationary distribution of flare fluxes, while individual years follow this model better. In three of the ten studied years, the data is consistent with a purely Poissonian quiescent distribution, while for five years only an upper limit of the flare flux distribution parameter could be determined. We find that these possible changes cannot be explained fully by the different number of observations per year. Combined, these results are instead consistent with a changing flaring rate of \sgra, appearing more active between 2006--2007 and 2017--2019, than between 2008--2012. Finally, we discuss this result in the context of flare models and the passing of gaseous objects, and discuss the extra statistical steps taken, for instance to deal with the background in the Swift observations.
\end{abstract}

\begin{keywords}
black hole physics -- Galaxy: centre -- X-rays: individual (\sgra) , stars: black holes
\end{keywords}

%%%%%%%%%%%%%%%%%
% INTRODUCTION
%%%%%%%%%%%%%%%%%

\section{Introduction}\label{sec:introduction}
Sagittarius A* (\sgra) is the electromagnetic counterpart of the supermassive black hole at the centre of the Milky Way galaxy. It has an estimated mass of $\sim 4\times 10^{6}~\Msun$, but its bolometric luminosity is $\sim 9$ orders of magnitude fainter than the Eddington luminosity for an object of this mass \citep{genzel2010galactic, morris2012}. It is the most nearby galactic nucleus, located at a distance of $\sim$ 8 kpc from Earth \citep{reid2004proper, ghez2008}, this makes \sgra\ the prime laboratory to study the accretion processes onto supermassive black holes at such low accretion rates.

The X-ray emission from \sgra\ is observed to be composed of a quiescent component, corresponding to a luminosity of $L_X \simeq 3\times 10^{33} \lum$ in the $2-10$~keV energy range, which is interrupted $\sim$daily by flares \citep[e.g.][]{baganoff2001, goldwurm2003, genzel2010galactic, markoff2010, neilsen2013, degenaar2013_sgra}. These flares are $\sim 1-2$ orders of magnitude more luminous than its quiescent emission, with the brightest ones reaching values of $L_X \simeq (1-5)\times 10^{35} \lum$ \citep[e.g.][]{nowak2012,haggard2019}. \sgra\ is also flaring in other wavebands, most prominently in the near-infrared \citep[nIR; e.g.][]{witzel2018}. Both, the emission mechanism and the physical process producing \sgra's flares are not completely understood yet \citep{markoff2001, liu2002, yuan2003, liu2004, cadez2008}, although \cite{ponti2017} developed the first simultaneous multiwaveband campaign measuring the spectral index in nIR and X-ray bands, showing that synchrotron emission with a cooling break is a viable process for \sgra’s flaring emission.

In the past decade, extensive work has been performed to simultaneously detect flares from \sgra\ in different wavebands, to gain more insight into the emission mechanism \citep{eckart2004, yusefzadeh2008, trap2011}. Another avenue of study has been to characterize the brightness distribution and occurrence rate of flares at X-ray and nIR wavelengths. In the X-rays, the first systematical work was developed before the passage of the G2 object to the Galactic Centre \citep[for reviews of the G2 object, see][]{gillessen2012, witzel2014}. Using 3 Ms of data from the \chan\ \textit{X-ray Observatory's} 2012 X-ray Visionary Project (XVP), \cite{neilsen2013} reported 39 flares for this set of data and a flaring rate of $\sim 1.1$ flares per day, which is consistent with previous results by \cite{genzel2010galactic}. It also has been shown that \sgra\ X-ray flux distribution can be decomposed into the sum of two processes: a quiescent component with constant flux, and a flaring component best described by a power law distribution of fluxes \citep{neilsen2015}. Regarding the nIR band, several works have also been performed. For instance, by making use of the GRAVITY instrument, \cite{abuter2018} detected orbital motions of three NIR flares from \sgra. Also, using data from the \keck\ Observatory, {\it Spitzer} Space Telescope, and the {\it Very Large Telescope (VLT)}, \cite{witzel2018} found that the variability of \sgra\ in this band can be described as a red noise process with a single log-normal distribution.

Establishing if the flaring properties of \sgra\ change over time could provide new hints into the physical mechanism producing the flares. For instance, \cite{ponti2015} claimed evidence for a change in \sgra\ flaring rate, using 6.9 Ms of \chan\ and \xmm\ data, covering September 1999 to November 2014. They reported a significant increase in the rate of the very bright flares at the late 2013 and 2014, changing from $0.27 \pm 0.04$ to $2.5 \pm 1.0$ flares per day at a 99.9\% confidence level. On the other hand, \cite{bouffard2019} studied a total of 4.5 Ms of \chan\ observations only, covering 2012--2018 data free of contamination from the magnetar SGR J1745-2900 (see section \ref{subsec:swiftlc}) but did not find evidence of change in \sgra\ flaring properties between the XVP and post-XVP data. Similarly, based on 4.5 Ms of \chan\ observations from 1999 to 2012, \cite{yuan2015} and \cite{yuan2017} did not find evidence of changes in the quiescent nor flaring rate of \sgra, even around the pericentre passage of the S2 star in 2002. Another recent work suggests the emission of \sgra\ in the nIR band has been consistent during $\sim 20$ year of observations \citep{chen2019}, but apparently the emission has reached unprecedented flux levels in 2019, with flux peaks that are twice the values from previous measurements \citep{do2019}.

In this work, we study the Cumulative Distribution Function (CDF) of count rates of \sgra\ using data accumulated between 2006 and 2019 with the Neil Gehrels \swift\ observatory  \citep[\swift;][]{gehrels2004}. The long-term monitoring and high observing cadence of the \swift\ program uniquely allow us to test whether the properties of \sgra's X-ray flaring behaviour show evidence of changes on a timescale of years.

%%%%%%%%%%%%%%%%%
% OBSERVATIONS/METHODS
%%%%%%%%%%%%%%%%%

\begin{figure*}
    \centering
    \includegraphics[width=\textwidth]{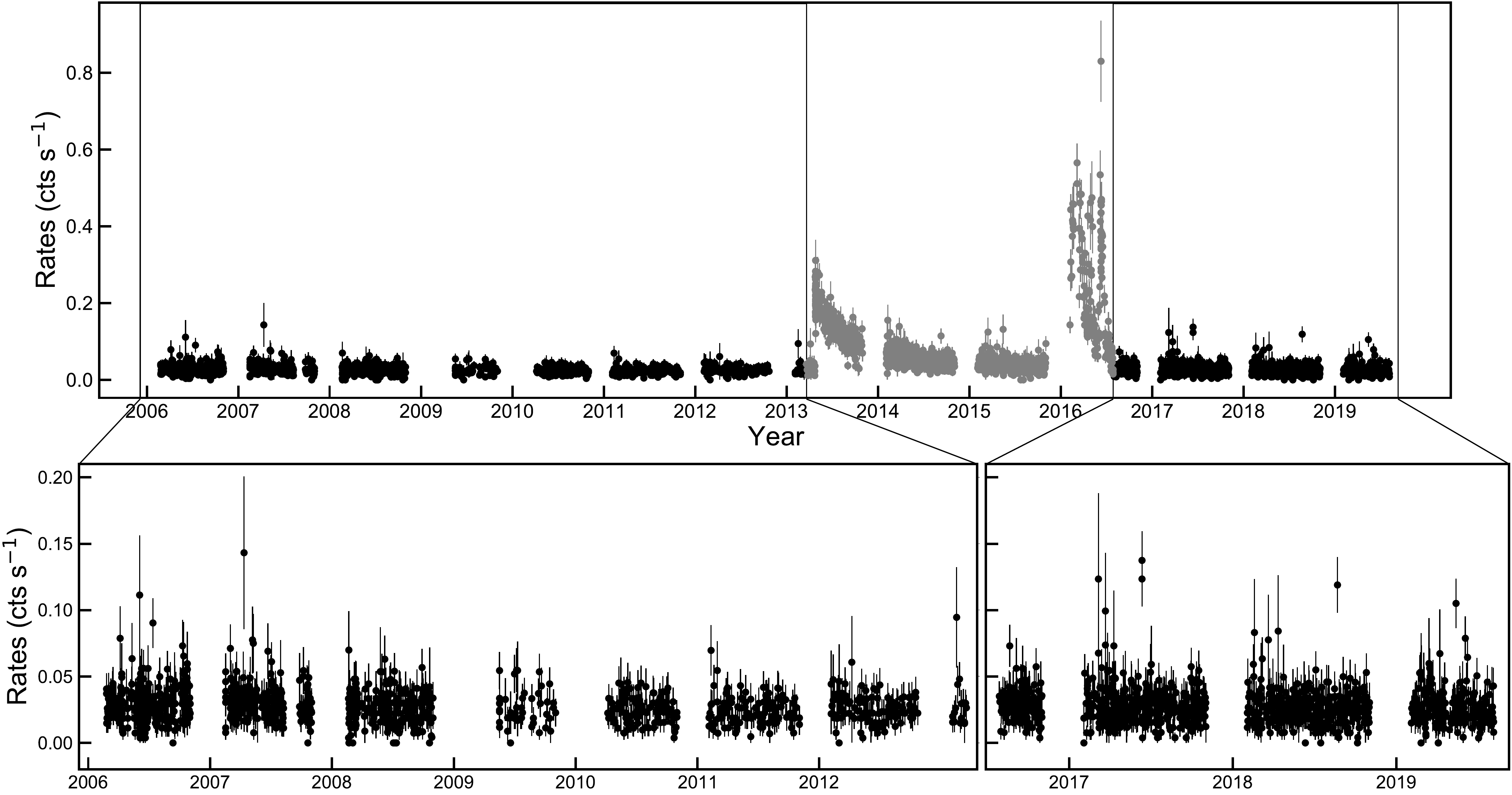}
    \caption{Long-term \swift/XRT-PC light curve of \sgra\ at 500-s binning (0.3--10 keV). The data marked in grey in top panel were excluded from our analysis due to the activity of nearby transient X-ray sources (see Section~\ref{subsec:swiftlc}). The bottom panel shows the \sgra\ data that we used for the analysis.} 
    \label{fig_lc}
\end{figure*}

\section{Observations and methods}\label{sec:methods}

\subsection{\swift/XRT long-term light curve}\label{subsec:swiftlc}
In 2006 February, \swift\ started to monitor the Galactic Centre (GC) with the on-board X-Ray Telescope (XRT) with the aim of studying \sgra\ as well as numerous transient X-ray binaries located in this region \citep[][]{kennea2006}. Apart from a handful of interruptions and Sun constraints, \swift/XRT has pointed at the GC every $\sim$1--3 days since 2006, with an average exposure time of 1~ks per observation \citep[see][for a review of the program]{degenaar2015_swiftreview}.

In this work we used all available \swift/XRT data that covered \sgra\ and was obtained in photon-counting (PC) mode. The data spans the period between 2006 February 24 and 2019 August 6. The light curve of \sgra\ was extracted with the software implemented in the online XRT data analysis tool \citep[][]{evans2007,evans2009}\footnote{https://www.swift.ac.uk/user$\_$objects/}, with the only exception that a fixed source and background region were used. To extract events from \sgra, we employed a circular region with a $10''$ radius centred at R.A. =  266.41682 and Dec. = -29.007797 (J2000). To account for the background, we used three circular regions of $10''$ that were free of X-ray point sources but did contain diffuse X-ray emission (as seen at the position of \sgra).

The long-term light curve was created with a bin size of 500~s and it is shown in Figure \ref{fig_lc}. It clearly shows two bright, extended periods of activity that do not belong to \sgra, but to transient sources located within $10''$ of the supermassive black hole. The first is the transient magnetar SGR~J1745--29, which exhibited an outburst between 2013 and 2015 \citep[][]{kennea2013,cotizelati2017}, and the second is the transient X-ray binary Swift~J174540.7--290015 that was active in 2016 \citep[][]{reynolds2016,ponti2016}. Given the brightness of these objects and the small angular separation of these from \sgra, compared to the point spread function of \swift, it is impossible to extract reliable information on the brightness of \sgra\ during the time that these transients were active. Therefore, we excluded all data obtained between 2013 March 31 and 2016 July 28 from our analysis of \sgra. 

Apart from the obvious outbursts of the two transients above, the \swift\ light curve of \sgra\ shows some instances of elevated emission that has been ascribed to flares from \sgra. Based on the significance of these high points compared to the long-term average XRT count rate at the position of \sgra, several bright X-ray flares have been reported previously \citep[][]{degenaar2013_sgra,degenaar2015_swiftreview,degenaar2019,reynolds2018}.

\subsection{Modelling the count rate distribution}\label{subsec:cdf}
In this work, we focus on the count rate distribution, instead of individual flares. We followed a similar approach to the one described in \cite{neilsen2015} to analyse the CDF of \sgra's light curve, shown in the bottom panel of Figure \ref{fig:cdfall}. The empirical CDF is the fraction of rates greater than or equal to a given rate, and is defined by the equation
\begin{equation} 
    f(r) = \frac{1}{N}\sum_{i=1}^{N} \mathds{1}_{\{r_i \geq r\}},
\end{equation}
where $N$ is the number of time bins, $\mathds{1}$ is the indicator function\footnote{The indicator function, $\mathds 1$, is defined to be 1 when the condition is true, and 0 elsewhere} and $r_i$ are the count rates. With this definition, $f(r)$ runs from one to zero, allowing for an easier visualisation of changes at the high rate end on a logarithmic scale: $f(0) = 1$ per definition, as count rates are non-negative. $f(r_{\rm h}) = 1/N$, where $r_{\rm h}$ is the highest observed count rate in our data, while $f(r) \equiv 0$ for any $r>r_{\rm h}$.

Following the results from \citet{neilsen2015}, we assumed that the emission from \sgra\ is composed of a quiescent and a flaring emission component. We model these components as a constant and a power-law distribution in flux, respectively. We opted for the power law distribution for the flux of the flaring component because \cite{neilsen2015} found such a distribution provides the best description of 2012 \chan\ XVP data -- since the \textit{Chandra} observations have a higher sensitivity then Swift, we do not expect to see deviations from this power law model in the CDF of a similar-length light curve (i.e. $\sim$one year). To generate the flaring component, one can define a power law probability distribution of fluxes straightforwardly as

\begin{equation}
    \label{eq:pl}
P(F) = 
    \begin{cases}
    kF^{-\alpha}, & F_{\rm min} < F < F_{\rm max}\\
    0, & \text{elsewhere},
    \end{cases}
\end{equation}
where $F$ represents the flux, $\alpha$ the power law index and $k = (\alpha -1)/(F_{\min}^{1-\alpha} - F_{\rm max}^{1-\alpha})$ is a constant of normalisation\footnote{This definition of $k$ corrects a small typo in the original model from \cite{neilsen2015}}. 

This definition, introduced by \cite{neilsen2015}, carries a significant downside, however, for data where flaring is either not present or faint in comparison to the quiescent rate and its uncertainties: the parameter characterising the flaring rate, $\alpha$, is unbounded. Low values of $\alpha$, approaching zero, imply an average flare flux that is relatively high within the considered range $F_{\rm min}$ to $F_{\rm max}$. On the other hand, if \textit{no} flaring is present, $\alpha$ will asymptotically tend to infinity as the flare flux tends to a constant value of $F_{\rm min}$ -- combined, the quiescent and flaring components then form the equivalent of a single quiescent component. In reality, our analysis shows that this effect shows up already when $\alpha \approx 3$--$4$, which means that any $\alpha \geq 3$ will yield the same model fit and quality. As a result, for a light curve without (detectable) flaring, $\alpha$ will be unconstrained at high values. 

To counteract this issue, we instead introduce a re-parameterisation to our analysis, defined as 

\begin{equation}
    \zeta \equiv \frac{1}{\alpha}
\end{equation}

\noindent This new flaring parameter $\zeta$ is mathematically better constrained, with the lower bound $0$ corresponding to the case of no flaring and its upper bound set by the distribution of flares in the data -- arbitrarily high values of $\zeta$ are therefore not possible, as those correspond to ever-increasing flare fluxes. Moreover, $\zeta$ is more intuitively interpreted, with low values corresponding to low flare fluxes. We stress that this approach does not fundamentally differ from e.g. \cite{neilsen2015}; instead, it simply allows us to calculate a proper upper limit on the flaring parameter when no flares are observed, which is not possible with the unbounded $\alpha$-parameter\footnote{Also, our \textit{Swift} analysis shows that $\alpha$ does not follow a Gaussian distribution, that would be distorted through this conversion. As such, our analysis does not lose any advantages that would come from assuming Gaussian distributions, due to this change of parameters.}. 

Furthermore, we note that the simulated fluxes do not depend strictly on $k$, but only on $F_{\rm min}, F_{\rm max}$ and $\zeta$.  We fixed the values (in units of $10^{-12}~\flux$) to be $F_{\rm min} = 0.1$ and $F_{\rm max} = 16.0$ for the minimum and maximum flux, respectively. This $F_{\rm min}$ is higher than that from \citet{neilsen2015}, to account for the lower sensitivity of \swift\, and corresponds to its limiting sensitivity in 500~s. For the maximum, we took the brightest flare measured with \swift\, as reported by \cite{degenaar2013_sgra}. 

In the case that no significant background emission is present (as can be done for, e.g., the \textit{Chandra} XVP data), the flaring flux can simply be converted into a count rate assuming a certain flux-to-counts conversion and Poisson statistics. However, given the relatively high background in the \textit{Swift} Galactic Centre data, Poisson statistics do not hold. In the source region, we measure the sum of source counts and background counts, both of which are Poisson distributed. The former is then corrected by subtracting the background counts as measured in a separate region, after which the remaining source counts as modelled are the sum of quiescent and flaring emission. Assuming the background is spatially constant, this two-component model correctly describes the \textit{mean} of the source counts, since the mean of two subtracted Poisson variables equals the subtraction of their respective means. However, the resulting \textit{distribution} of source counts does not follow a Poisson but a Skellam distribution, defined by the total counts in the source region and those in the background region. 

\begin{figure}
    \centering
    \includegraphics[width=8.3cm, height=4.5cm]{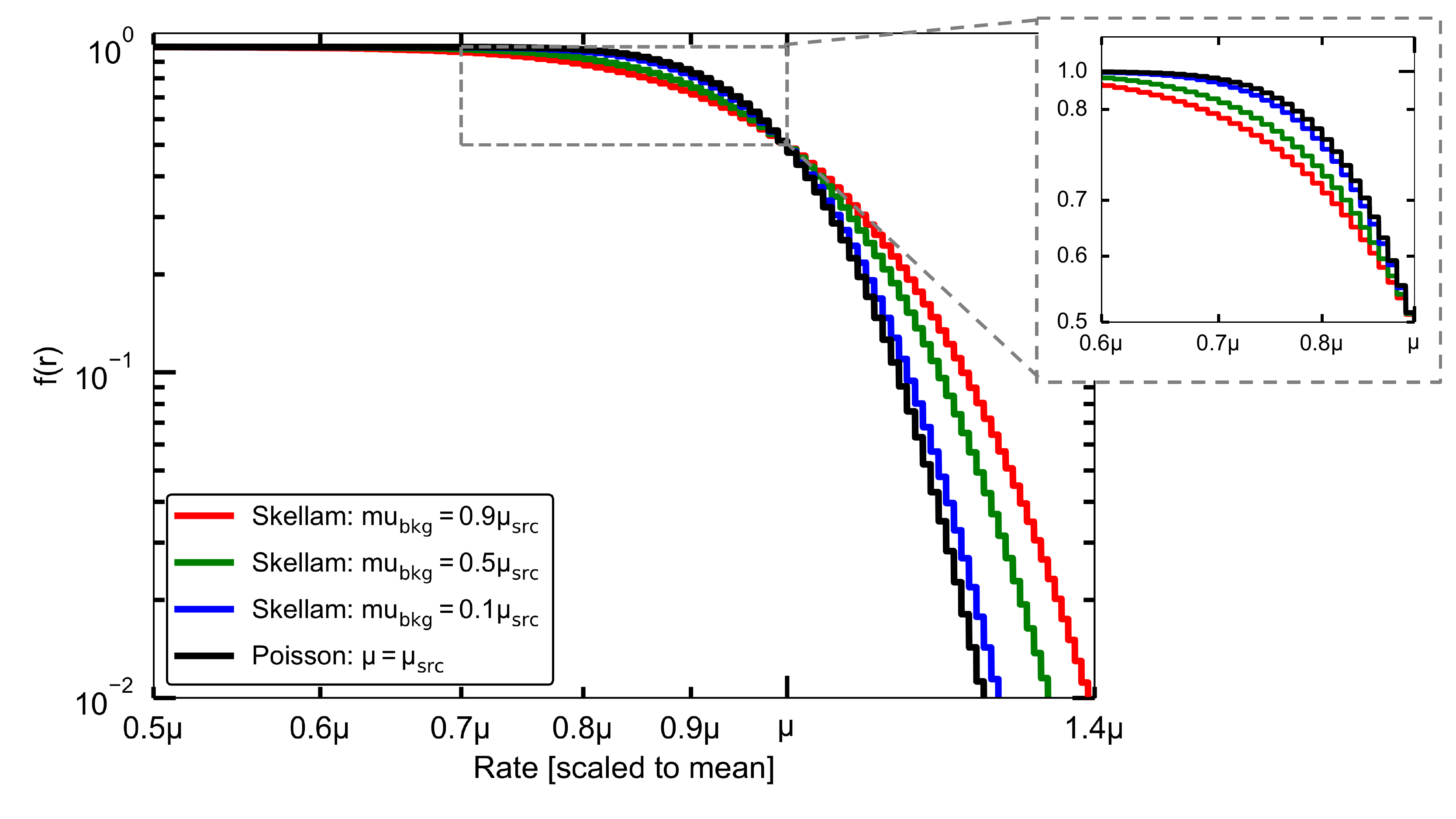}
    \caption{The CDF shape for one Poisson and three Skellam-distributed datasets with the same mean $\mu_{\rm src}$, assuming different relative background levels $\mu_{\rm bkg}$.}
    \label{fig:skellam} 
\end{figure}

To highlight the difference between the Poisson and Skellam distributions, we plot four examples of the CDF $f(r)$ in Figure \ref{fig:skellam}. The three examples for Skellam distributions, with background rates at $90$, $50$, and $10$\% of the mean count rate, show the significant distortion to the CDF shape. In the case of our \textit{Swift} data, we find that the ratio $b/\mu \approx 0.1$, and hence we need to account for this distortion by explicitly modelling the data using the Skellam distribution. Otherwise, the CDF shape would be incorrectly reproduced and, possibly, the flaring flux would be overestimated. 

Therefore, we model the distribution of source count rates, assuming a value of $\zeta$ and the quiescent count rate $Q$, using the equation

\begin{equation}
\label{eq:plflux}
    r(F, Q) = \frac{1}{\Delta t}\mathtt{skellam}\left[\left(F\times\left(\frac{r_{\rm max}}{F_{\rm max}}\right) + Q + b\right)\Delta t, b\Delta t\right],
\end{equation}

\noindent where $\Delta t = 500$~s is the bin size we used, $\mathtt{Skellam}$ is a Skellam random number generator, $b$ is the measured background region count rate, $F$ is the simulated flux taken from Eq \eqref{eq:pl}, and $r_{\rm max} = 10.2 \times 10^{-2}$~cts/s is the mean observed count rate of the brightest flare \citep{degenaar2013_sgra}. The Skellam random number generator takes the total source and background region counts as its two inputs, and returns a random draw of the former corrected by the latter. We applied a constant-rate conversion because no significant variations in the spectrum of \sgra\ flares have been observed \citep[e.g][]{degenaar2015_swiftreview}. Furthermore, we are not applying a pile-up correction, because the count rates seen by \swift, even in flares, from \sgra\ are all below 0.2 ct/s, i.e. well below the rates where this effect becomes important (see Figure \ref{fig_lc}).

To fit the observed X-ray distribution of \sgra, we generated sets of synthetic data consisting of a combination of: a quiescent count rate distribution, with mean $Q$, and  a power law flux distribution of index $\zeta \equiv 1/\alpha$. To find the values of the parameters that best match the observations, we considered the two methods explained below.

\subsubsection*{Two dimensional method}
\label{sec:2D}
Following the same approach as \citet{neilsen2015}, in the two-dimensional (2D) method, we create a grid for values of $\zeta$ between $0$ and $1$,  and values of $Q$ between $14$ cts/ks and $26$ cts/ks, with 100 steps in both. Fits on larger grids show that outside these ranges, the probability of a match between the real data and synthetic data is null and/or unchanging. 

For every pair $(\zeta, Q)$, we generated 1000 sets of synthetic distributions and applied the Anderson-Darling test \citep{anderson1954} (hereafter called AD-test) to each combination of the real and a synthetic light curve to compare their distributions. At each parameter pair, we stored the average Test-Statistics (TS) from the AD-tests and then we converted it to p$-$values to assess the probability of both distributions to be drawn from the same underlying distribution. We then computed the median value and the $1-\sigma$ confidence interval of $\zeta$ and $Q$ from the marginalised probability contours generated in the $\zeta - Q$ plane; alternatively, if the distribution did not converge to zero at low $\zeta$, we instead calculated the $90$\% upper limit of $\zeta$. We note that the measured values of $\zeta$ and $Q$ depend on $F_{\rm min}$, since for larger values of $F_{\rm min}$, the calculated value of $\zeta$ decreases, while $Q$ decreases (see Discussion). However, as we use a single value of $F_{\rm min}$ in the entire analysis, this does not affect any differences seen between subsets of the data. We used the \textsc{python}/\textsc{scipy} function \textsc{scipy.stats.anderson\_ksamp} to compute the AD-test; as this function is only designed to convert the AD test statistics (TS) to p-values between $p=0.001$ and $p=0.25$, we supplemented this conversion with Monte-Carlo simulations as detailed in Appendix \ref{appendixA}.

In the AD-test comparison between the synthetic and observed count rates, the time-dependence of these quantities is ignored. However, given that flares may last longer than the choosen time bin size ($500$ sec), the count rates in consecutive time bins are not necessarily independent \citep{li2015}. In our analysis, we will therefore observe a steeper flare flux distribution (i.e. a relatively higher number of faint flares), as the total flare fluence is divided over multiple time bins. This paper is mainly focused on the comparison between sub-sets of the \textit{Swift} campaign, where the same steepening of the distribution is expected. Therefore, we argue that sub-sets of the data can accurately be compared with this method.

Before moving to the one-dimensional method below, we make a brief note regarding the above method to calculate confidence intervals. In addition to the method described above, \citet{neilsen2015} also applied an MCMC fitting routine to directly calculate and maximise the likelihood from the \textit{Chandra} light curve. Comparing these two methods, one finds that both approaches return consistent results for the best fit parameters. However, the AD-test method returns slightly more conservative estimates of the confidence levels. Due to its computational speed, allowing us to perform a variety of tests on the long-term \textit{Swift} data, we opted to use the AD-test method. However, aiming to remain conservative in our comparison between different sub-sets of the \textit{Swift} data (see Section \ref{sec:results}), we did not correct for a possible over-estimation of confidence levels.

\subsubsection*{One dimensional method}
The above method assumes that $Q$ and $\zeta$ are independent, while not all combinations will be able to represent the real light curve: for instance, a high $Q$ will require a small $\zeta$ to prevent systemtically overpredicting the light curve rates. Therefore, as a complementary one dimensional (1D) method, we again assumed each point in our light curve consists of the sum of a quiescent and a flaring component, both with Poisson noise. Therefore, for each datapoint, we can state that the rate $r_i$ is given by $r_i = (1/\Delta t)\left[ \mathtt{prand}(Q\Delta t) + \mathtt{prand}(F\times (r_{\rm max}/F_{\rm max})\Delta t)\right]$. Since the sum of two Poisson random variables with mean $\lambda_1$ and $\lambda_2$ is again a Poisson random variable with mean $\lambda_1 + \lambda_2$, and the fact that $\Delta t$ is a constant, then the mean of all the observations in our data set will be the sum of the mean of both components: $\overline{r} = \overline{Q} + \overline{F}\times (r_{\rm max}/F_{\rm max})$\footnote{Here, we ignore the background region count rate as we assume that the background is spatially constant. The introduction of the Skellam statistics is therefore only necessary when considering the distribution of subtracted data, not the mean.}. Computing $\overline{F}$ from Eq. \eqref{eq:pl}, we can express the mean count rate for our light curve as

\begin{equation}
\label{eq:1d}
\overline{r} = \overline{Q} + \frac{r_{\rm max}}{F_{\rm max}}\left(\frac{1-1/\zeta}{2 - 1/\zeta} \right)\frac{F_{\rm max}^{2-1/\zeta} - F_{\rm min}^{2-1/\zeta}}{F_{\rm max}^{1-1/\zeta} - F_{\rm min}^{1-1/\zeta}},
\end{equation}
where the second term of the right-hand side of Eq. \eqref{eq:1d} corresponds to mean value of fluxes, $\overline{F}$.

Therefore, we calculated the mean count rate of our data, $\overline{r}$, and we assumed fixed values of $F_{\rm max}, F_{\rm min}$ and $r_{\rm max}$. For a given $\zeta$, we computed $\overline{F}$ and then we used Eq. \eqref{eq:1d} to calculate $\overline{Q}$. With those $\zeta$ and $\overline{Q}$, we generate 1000 sets of synthetic data, and stored the average TS from the AD-test. We repeated this process on a non-linear grid of $\zeta$ between $0$ and $1$ (focusing on values between $0.4$ and $0.8$ with higher resolution after inspection of the final PDF shapes for low-resolution grids). We then converted the average TS at each $\zeta$ into $p-$values and again we calculated the median value of $\zeta$ with its $1-\sigma$ confidence interval. 

This method has the advantage to be $\sim 100$ times quicker than the two dimensional method, so it allowed us to run more simulations and re-draw data in order to find significance of changes in our analysis (see Section \ref{sec:results}). The downside is that it assumes the mean count rate, $\overline{r}$, does not have an uncertainty; therefore, we also applied the 2D method to compare their results for the analysis of individual years. 

%%%%%%%%%%%%%%%%%
% RESULTS
%%%%%%%%%%%%%%%%%

\section{Results}
\label{sec:results}
\begin{figure}
    \centering
    \includegraphics[width=\columnwidth]{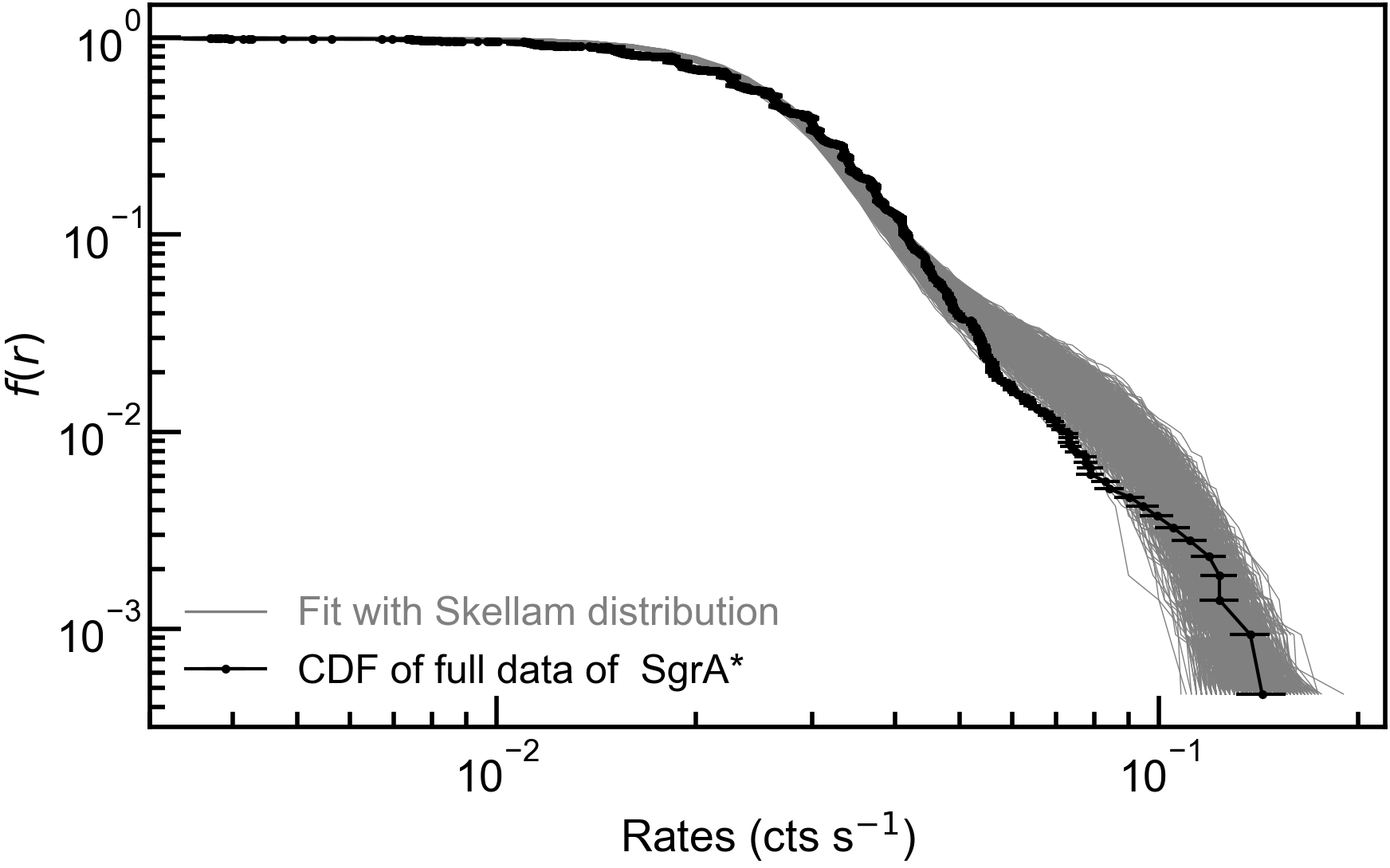}
    \caption{Empirical CDF of count rates of \sgra\ for the entire 2006--2019 \swift\ dataset (as described in section \ref{subsec:swiftlc}) is shown in black. The grey lines are 1000 sets of simulated data following a Skellam distribution. We note that in this figure, model components should be added vertically; likewise, the CDF uncertainties are vertical, as the fractions are set by the number of data points in the light curve.}
    \label{fig:cdfall} 
\end{figure}

%\subsection{CDF fitting results}
The empirical CDF of \sgra\ obtained for the entire 2006--2019 light curve is shown in black in Figure \ref{fig:cdfall}. We found that the quiescent component of \sgra\ emission is best represented by a pure Poisson distribution with mean count rate of $Q = 22.3_{-1.4}^{+1.6}$ cts/ks, and a power law process with $\zeta = 0.57_{-0.12}^{+0.18}$; for both parameters, throughout this paper, we will report the mode, with uncertainties corresponding to the $1\sigma$ level. These values of $\zeta$ and $Q$ were calculated with the 1D method, while the 2D method returns consistent values. The grey lines in Figure \ref{fig:cdfall} are 1000 different synthetic sets, being the combination of both distributions as established in Equation \ref{eq:plflux}. We can notice there is a clear model excess towards the tail of the distribution, especially between $\sim 0.05$ and $0.1$ counts per second, corresponding to the flaring component in the synthetic data. This discrepancy cannot be improved by changing the value of $\zeta$ nor $Q$ in the simulations. It also implies that the full 2006--2019 \swift\ data cannot be described by a stationary Poisson + power-law model as well as the \textit{Chandra} 2012 data in \cite{neilsen2015}, despite the higher sensitivity of \textit{Chandra}.

Since \cite{neilsen2015} showed this model can reproduce the 2012 \chan\ count rate distribution of \sgra, while it does not appear to work similarly well for all \swift\ data, time-variability in the CDF might play a role. To study whether this might be the case, we first divided the data into two sets and reanalysed the CDFs: set $A$, containing the five years with the highest maximum count rates measured by \swift\ (i.e, 2006 -- 2007 and 2017--2019), and set $B$, containing the five years with the lowest maximum count rates (2008 -- 2012). We only applied the 1D method for this, as testing the significance of any differences between the two sets required simulations that were too computationally expensive in the 2D method (see below). 

\begin{figure}
    \centering
    \includegraphics[width=\columnwidth]{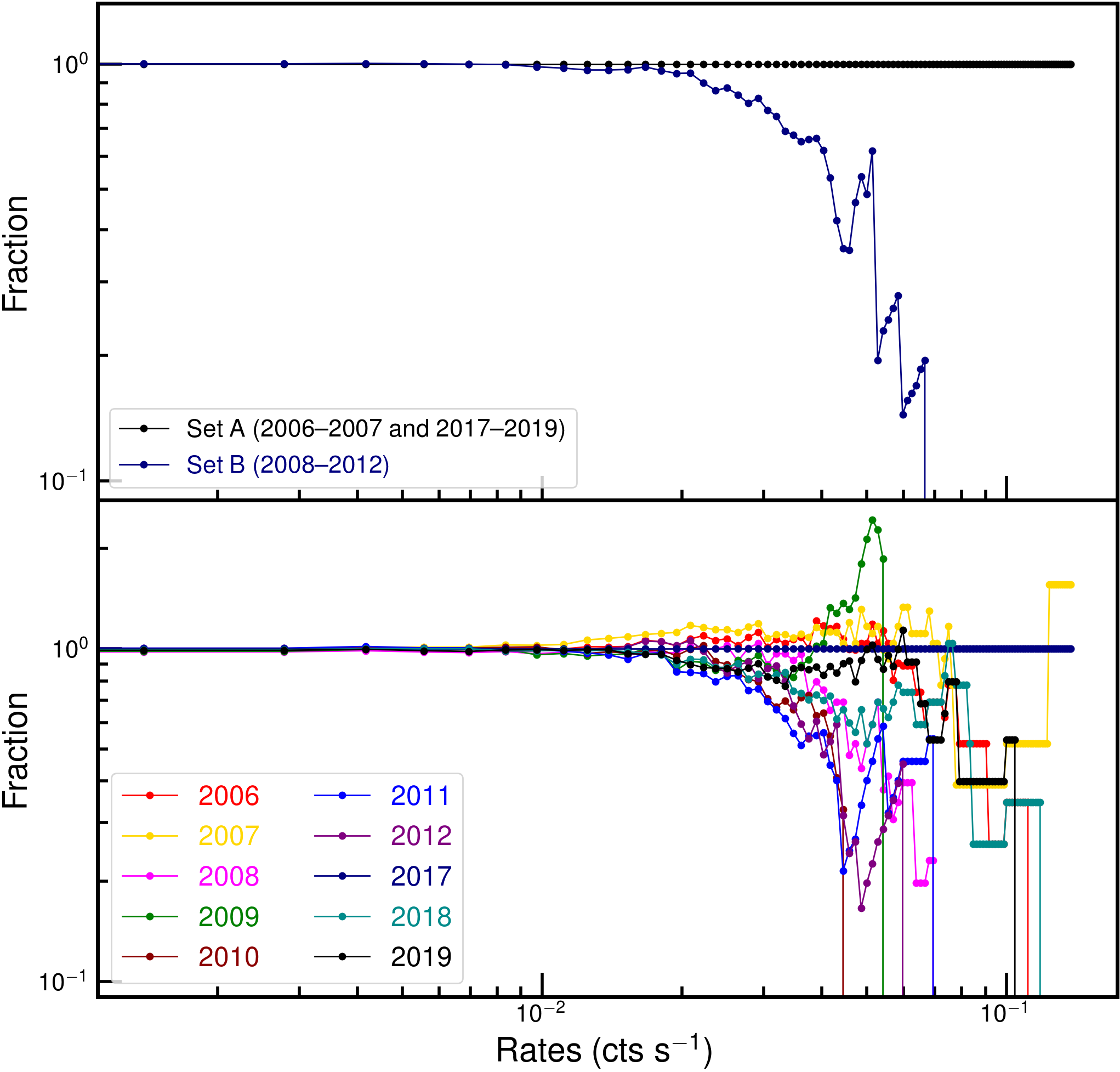}
    \caption{\textbf{Top panel:} fraction of the CDF for set $A$ (2006--2007 and 2017--2019) and set $B$ (2008--2012). The reference line in black is the fraction $f_{A}(r)/f_{A}(r)$ \textbf{Bottom panel:} Same as top panel, but the ratio is for individual years. The reference line in navy blue is the fraction $f_{2017}(r)/f_{2017}(r)$.}
    \label{fig:cdf_ind}
\end{figure}

\begin{figure*}
\includegraphics[width=\textwidth]{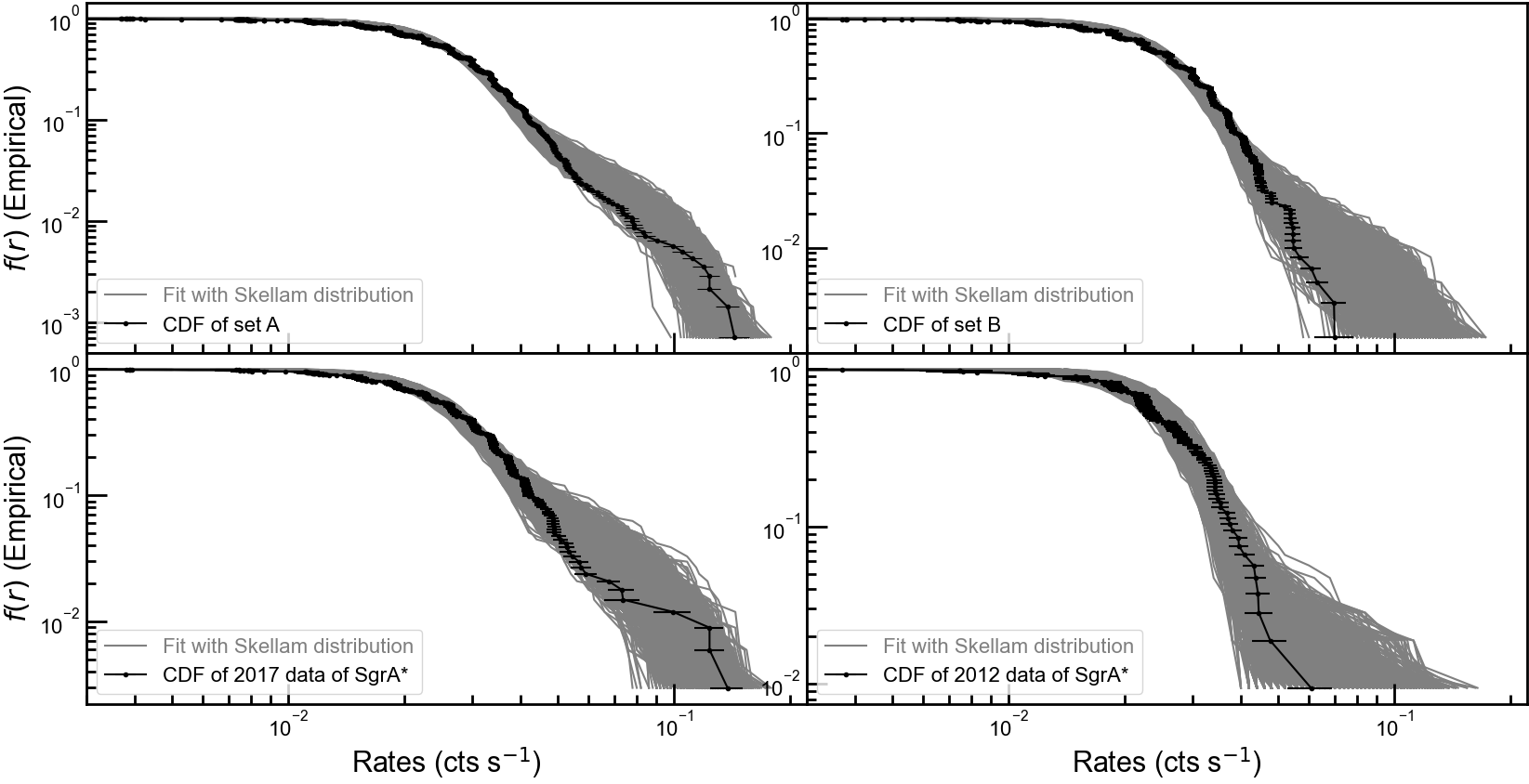}
\caption{Results of the fits for: \textbf{Top left panel:} Set $A$, consisting of the years with the highest count rates of \sgra; \textbf{Top right panel} set $B$, consisting of the years with the lowest count rates of \sgra; \textbf{Bottom left panel} 2017 data of \sgra; \textbf{Bottom right panel:} 2012 data of \sgra. All these fits correspond to the values of $\zeta$ and $Q$ calculated with the 1D method.}
\label{fig:cdf_ab}
\end{figure*}

The distinction between these sets can be seen in the top panel of Figure \ref{fig:cdf_ind}, where we plot the ratio of the CDF for set $A$ and set $B$, both with respect to the CDF of set $A$. The clear difference visible between the CDFs from sets $A$ and $B$ indeed hints toward changes in flaring rate over time: the CDFs of the two sets start to diverge around $\sim 0.02$ counts/second, much lower than the flare rates used to select set A and B. Therefore, we first tested, using the AD test, whether set A and set B are consistent with being drawn from the same distribution. We find $TS = 3.30$, corresponding to a p-value of $0.015$. The CDF modelling for these two sets is shown in the top-left and top-right panels of Figure \ref{fig:cdf_ab}: we can notice a visual improvement in these fits compared with the full set, particularly for set $A$; in set $B$, the flaring rate appears to be overestimated still in a majority of plotted synthetic light curves. The values of $\zeta$ and $Q$ that we measured for these sets are summarised in Table \ref{tab:test}. 

\begin{figure}
    \centering
    \includegraphics[width=\columnwidth]{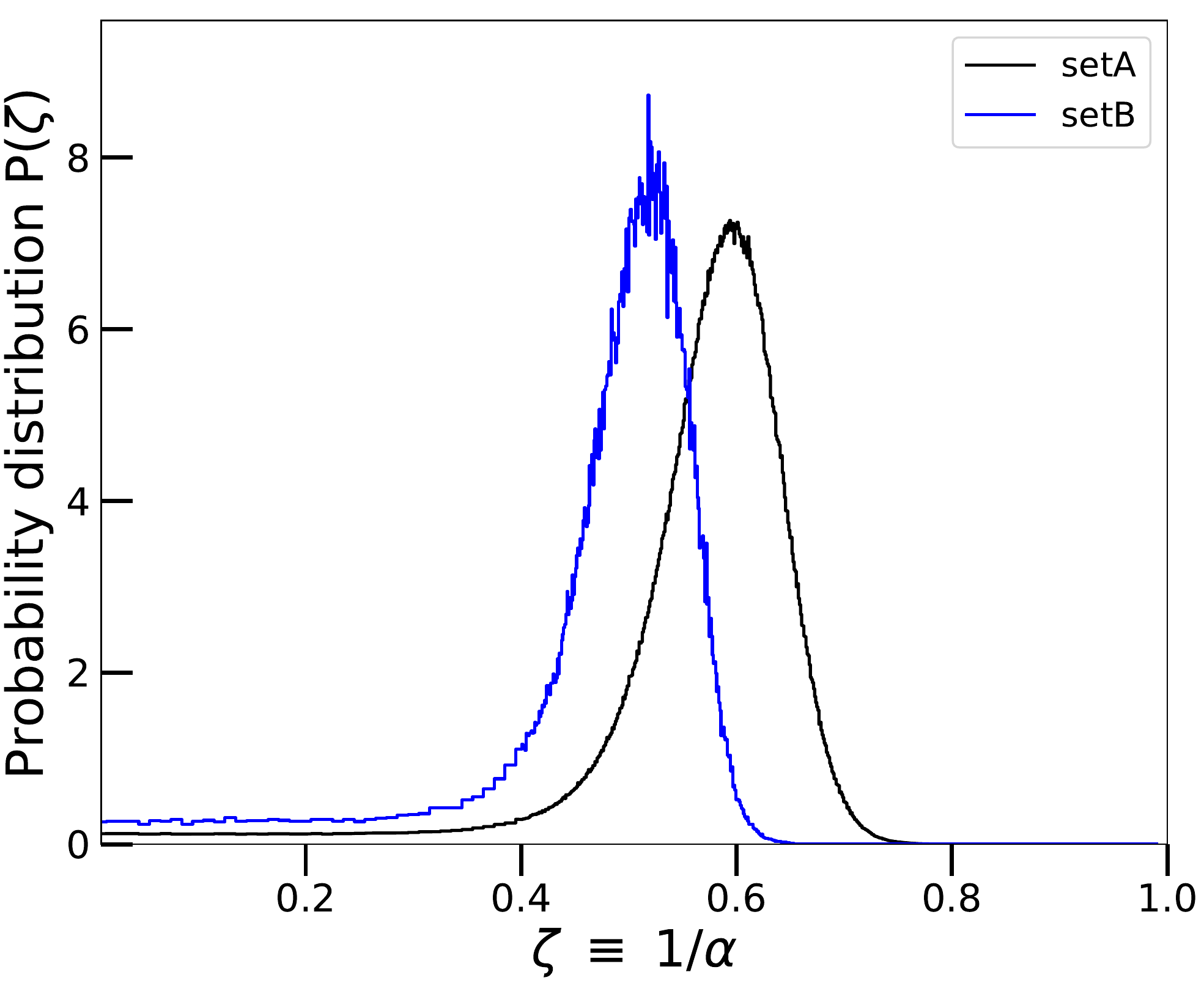}\\
    \includegraphics[width=\columnwidth]{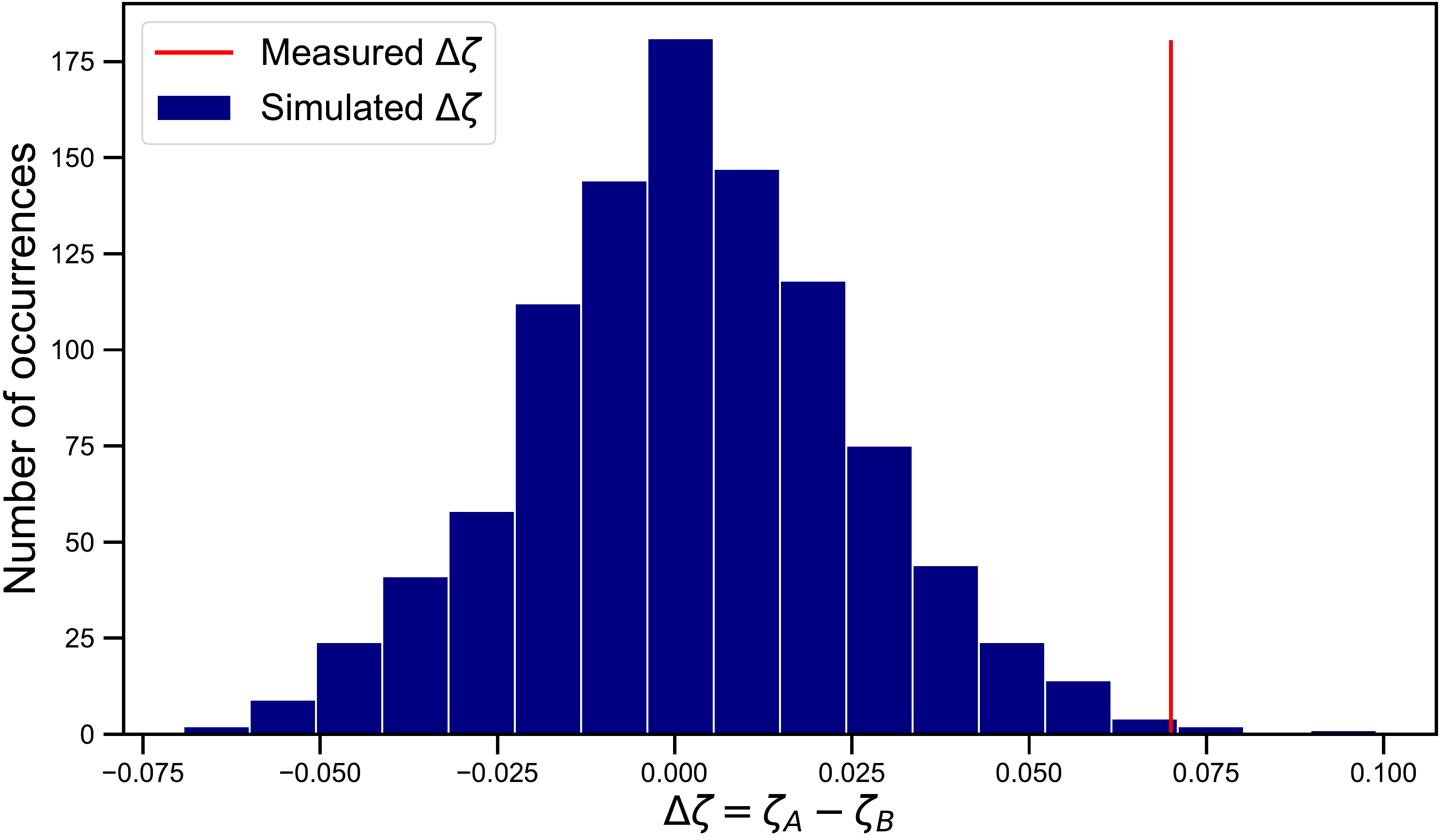}
    \caption{\textbf{Top panel:} Probability distribution as function of $\zeta$ for set A and set B. The maximum in the graphs gives the best $\zeta$ for each set. \textbf{Bottom panel:} Histogram of occurrences of $\Delta \zeta$ in randomized data of set $A$ and set $B$ using a fix value of $F_{\rm min} = 0.10$. The measured value of $\Delta \zeta$ is shown in red. Most of the simulations have a $\Delta \zeta \sim 0.0$, while only three simulations show $|\Delta \zeta| > 0.07$.}
    \label{fig:histogram}
\end{figure}

The possible presence of a difference in the flaring component between set $A$ and $B$, translates into a difference in fitted power law index, although both values are consistent within their $1-\sigma$ intervals -- see also their probability density functions of $\zeta$ in the top panel of Figure \ref{fig:histogram}. We measured the difference in power law index with the parameter $\Delta \zeta = \zeta_A - \zeta_B = 0.07$. However, this difference and the apparently better fits for the two sets compared to the complete light curve, might simply be due to the smaller number of data points. Therefore we tested the hypothesis that such a change could arise at random when splitting the data in two, in the following way: we created two sets with the same number of data points of set $A$ and set $B$, but the count rates were taken randomly from the entire set of data. We then fitted those sets and calculated the value of $\Delta \zeta$ and repeated the process 1000 times. This test was performed by applying the 1D method, taking advantage of its faster performance. The histogram of occurrences for $\Delta \zeta$ is shown in the bottom panel of Figure \ref{fig:histogram}. We can observe that $|\Delta \zeta| \geq 0.07$ occurs merely three times in the randomized data. Therefore, the probability that a change in flaring distribution of this magnitude arises by chance, \textit{only} due to the decrease in data points, is $\sim 0.3\%$. We have repeated this test for different values of $F_{\rm min}$ ($F_{\rm min} = \{0.05, 0.1, 0.2\}$), redoing both, the fit of the real data and creating 1000 synthetic datasets; the same conclusion can be drawn in those cases.

With these results, we decided to further investigate the possible variability by dividing the data into individual years. First, we used a 1-sample AD-test, to test whether the data for the individual years is inconsistent with a pure Poisson process. We find that for the years 2010, 2011 and 2012, we cannot reject the hypothesis that a pure Poisson process underlies their light curve at $p<0.01$: their p-values are 0.03, 0.01, and 0.05, respectively. For the other years, we do find $p<0.01$, indicating the significant presence of a flaring component. 

\begin{figure}
    \centering
    \includegraphics[width=\columnwidth]{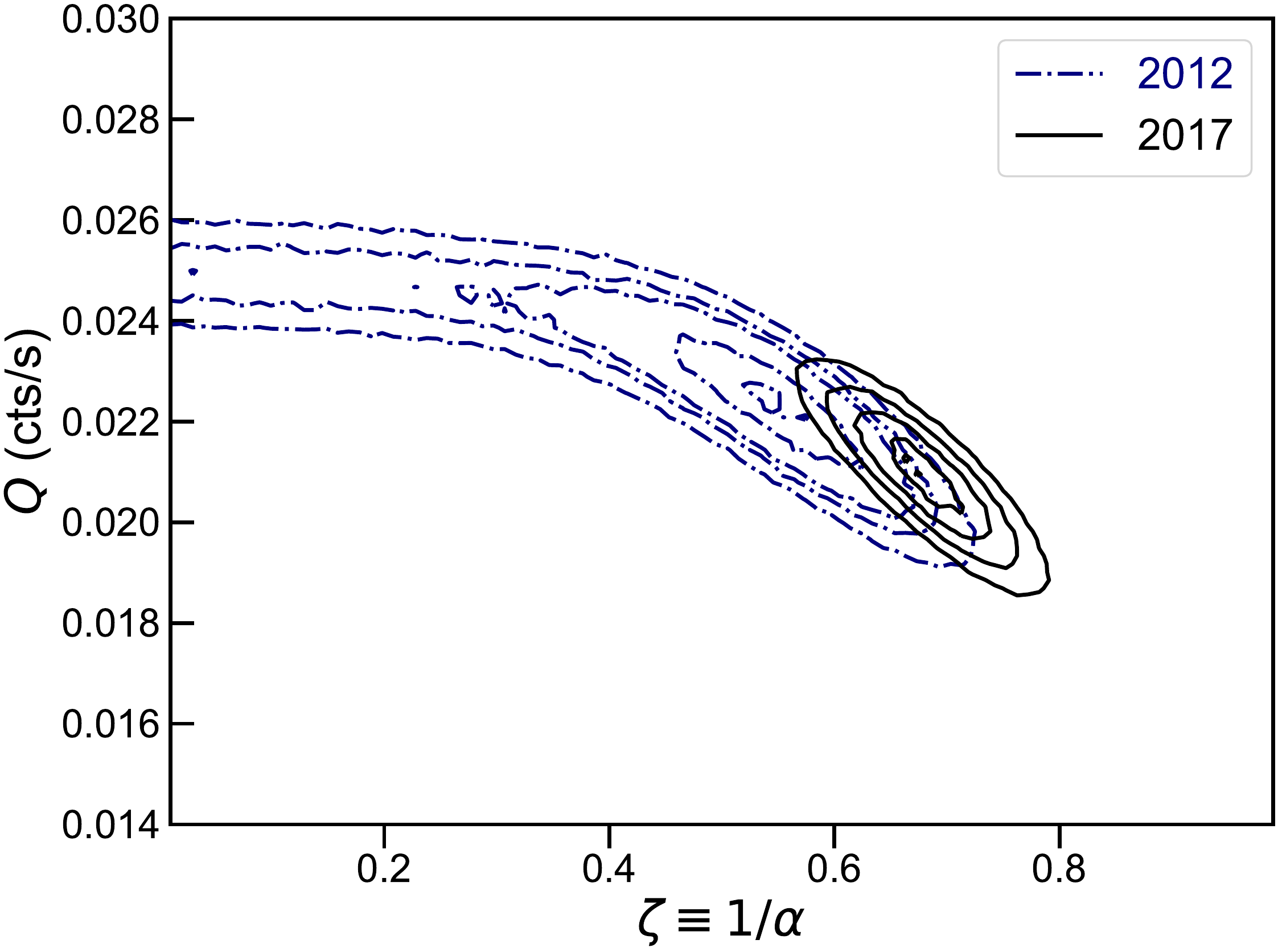}\\
    \includegraphics[width=\columnwidth]{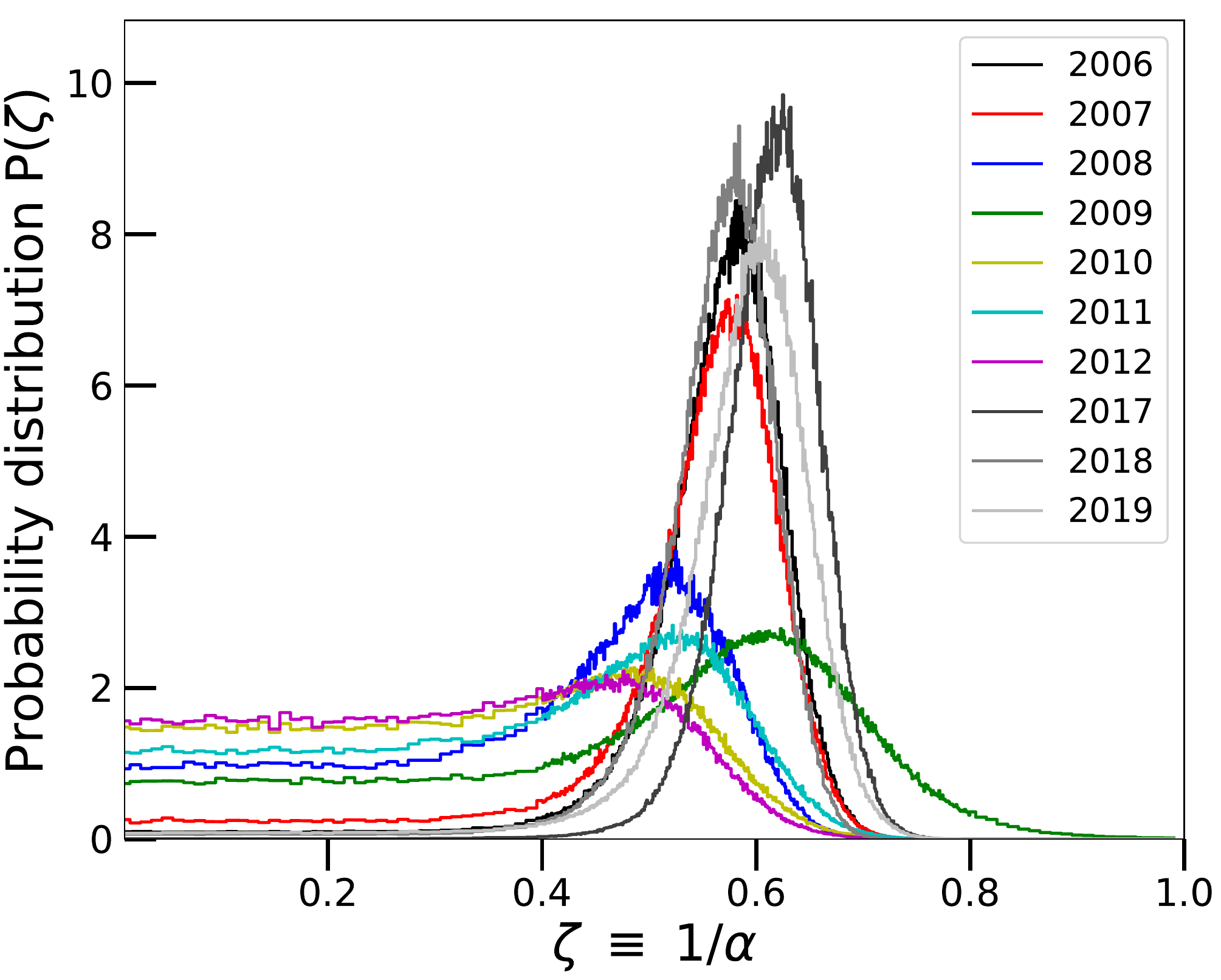}
    \caption{\textbf{Top panel:} Probability contour plots in the $\zeta-Q$ plane for 2017 data (black lines), 2012 data  (blue dashes). \textbf{Bottom panel:} Probability density from the 1D method, as function of $\zeta$ for 2017 and 2012 data.}
    \label{fig:contour_plots}
\end{figure}

Then, we turned to analysing the CDFs of the individual years. The results are summarised in Table \ref{tab:test}. We find that for several years, only an upper limit on the flaring parameter $\zeta$ can be measured. In those cases, the (marginalized) probability density function of $\zeta$ does not tend to zero as $\zeta \rightarrow 0$, and we list the $90$\% upper limit on the $\zeta$ instead of the mode. We observe this effect in 2009--2012 for both methods and 2008 in the 1D method only. In the 2007 data, an upper limit is obtained from the 2D method, while for the 1D method, the probability as $\zeta \rightarrow 0$ is close to, but not exactly, zero. For this borderline case, we therefore list the mode with uncertainties in Table \ref{tab:test}, but note that its $90\%$ upper limit would be $\zeta < 0.63$. We can see this comparison of individual years graphically in Figure \ref{fig:contour_plots}: the top panel shows how the $\zeta$-$Q$ contour plots from the 2D method do not close for the 2012 data, while they do for the 2017 data. While overlapping, the probability contour of 2012 data is systematically shifted to the left compared with the one for 2017 data. In the bottom panel of Figure \ref{fig:contour_plots}, we show the probability density, $p(\zeta)$ from the 1D method as function of $\zeta$. A clear maximum is again observed for 2017 data, but not for 2012 data, where after reaching the maximum, $p(\zeta)$ remain almost constant, indicating that for lower $\zeta$, fits of similar quality are obtained. It is for those years that show similar behaviour, that we calculate a $90$\% upper limit on $\zeta$ (in no cases do we find that the resulting probability interval of $Q$ is unclosed).

Splitting up the data, one can wonder whether the inability to constrain $\zeta$ in several years is simply due to a lower observing cadence: from 2009--2012, the \swift\ cadence was only one observation per 3 days, and in 2009 there were no observations between February and April \citep{degenaar2013_sgra}. To test the effect of a lower cadence, we simulated synthetic data sets with a $\zeta$ and $Q$ similar to those found in 2006--2007 or 2017--2019, but with the low cadence of the years between 2009--2012. When we reapplied the 2D method to these synthetic low-cadence years, we do find closed contour plots for $\zeta$ between $0.01$ and $1.00$ with the 2D method. A second test is to look more closely at Table \ref{tab:test}: notably, in 2019 contained a smaller number of datapoints (210) than 2007 (215) and 2008 (242); however, in 2019, $\zeta$ can clearly be measured, while only an upper limit could be measured for 2008 (1D-method) and 2007 (2D-method), and 2007 is also a borderline case in the 1D method. 

Therefore, we conclude that a lower cadence can contribute to poorer constraints on $\zeta$, for instance in 2009 (with only 49 points), but cannot account for all differences between years. This behaviour, together with the inability of a single flare distribution to reproduce all 2006--2019 observations, and the unconstrained $\zeta$ parameter between 2009 and 2012, can be interpreted as additional evidence for a change in the X-ray flaring emission properties from \sgra. We show all other contour maps, CDF fits and probability density functions of $\zeta$ for individual years in Appendix \ref{appendixB}. 

{ % begin box to localize effect of arraystretch change
\renewcommand{\arraystretch}{1.5}
\begin{table}
\centering

\caption{Measured values for the quiescent ($Q$) and flaring ($\zeta$) parameters found for different data sets of \sgra\ data. As the best-fit value we report the mode of the distribution, while uncertainties are quoted at 1 sigma and upper limits at $90$\%. Set A consists of 2006, 2007, 2017, 2018, 2019, while the rest of the years comprise Set B.} 
\label{tab:test}
\begin{threeparttable}
    \begin{tabular*}{\columnwidth}{l|ccccc} \hline
        Year & $\zeta$ (2D) & $Q$ (2D) & $\zeta$ (1D) & $Q$ (1D) & N  \\
         & & (cts/ks) & & (cts/ks) & \\ \hline
        Set $A$ & & & $0.59_{-0.08}^{+0.04}$ & $22.1_{-1.7}^{+2.0}$ & 1406 \\
        Set $B$ & & & $0.52_{-0.09}^{+0.03}$ & $22.1_{-1.2}^{+1.5}$ & 603 \\ \hline
        
        2006 & $0.65_{-0.09}^{+0.08} $ & $21.4 _{-1.7}^{+1.6} $ & $0.59_{-0.08}^{+0.03}$ & $23.2_{-1.6}^{+1.9}$ & 323 \\
        2007 & $<0.71$ & $19.5 _{-1.4}^{+2.6} $ & $0.58_{-0.12}^{+0.03}$ & $24.9_{-1.9}^{+2.4}$ & 215 \\
        2008 & $0.64_{-0.21}^{+0.07} $ & $22.4 _{-3.0}^{+1.6} $ & $<0.57$ & $23.7_{-2.3}^{+1.6}$ & 242 \\
        
        2009 & $<0.85$ & $22.4_{-2.5}^{+4.5}$ &  $<0.70$ & $22.0_{-4.4}^{+3.8}$ & 49 \\
        
        2010 & $<0.70$ & $20.6_{-1.1}^{+3.5}$ & $<0.54$ & $23.0_{-2.2}^{+1.8}$ & 102 \\
        2011 & $<0.73$ & $21.8_{-1.0}^{+4.0}$ & $<0.58$ & $21.8_{-2.6}^{+2.0} $ & 104\\
        2012 & $<0.68$ & $19.8_{-1.0}^{+3.3}$ & $<0.53$ & $23.9_{-2.2}^{+1.5}$ & 106 \\
        
        2017 & $0.69_{-0.09}^{+0.06}$ & $23.1 _{-1.7}^{+1.5} $ & $0.62_{-0.05}^{+0.04}$ & $21.5_{-1.7}^{+1.7}$ & 335 \\
        2018 & $0.65_{-0.09}^{+0.06} $ & $24.2 _{-1.7}^{+1.3} $ & $0.58_{-0.06}^{+0.03}$ & $20.7_{-1.6}^{+1.8}$ & 323  \\
        2019 & $0.67_{-0.11}^{+0.08} $ & $24.5.5_{-2.1}^{+1.6} $ & $0.61_{-0.07}^{+0.03}$ & $20.3_{-1.8}^{+2.0}$ & 210 \\ \hline
    \end{tabular*}
\end{threeparttable}   
\end{table}
}

%%%%%%%%%%%%%%%%%
% DISCUSSION
%%%%%%%%%%%%%%%%%

\section{Discussion}\label{sec:discuss}

The fits presented in this work of the CDF of \sgra\ count rates measured with \swift, show that it is not possible to adequately describe the complete set of X-ray observations using the model of a single power law process combined with a pure Poisson process, as shown in Figure \ref{fig:cdfall}. The excess in the tail of the distribution is not due to the power law index only, but also due to the high Poisson mean rate used in the quiescent component. Creating synthetic datasets with less flaring activity (i.e, lower $\zeta$) would require a higher value of $Q$ in the quiescent component, and then the lower count rates will be overestimated. Nevertheless, the emission can be described with this model better once we divide the data in different sets, as shown in Figure \ref{fig:cdf_ind}. 

Looking at the values from the 1D method in Table \ref{tab:test}, we notice that the flaring properties of the years forming set $A$ appear similar: the difference in best-fit $\zeta$ is small in those years, ranging from 0.62 to 0.58. On the other hand, for the years forming set $B$, the 1D-approach returns similar upper limits on $\zeta$, with the exception of 2009; these upper limits are typically below the best-fit flaring parameters for the years in set $A$, but do overlap with their uncertainty intervals. This result may suggest that the issues in fitting the full set of \swift\ data are not necessarily due to an incorrect model -- as expected given the adequate description of the \textit{Chandra} XVP data with this model \citep{neilsen2015}. Instead, it could also fit with the idea that the flaring rate of \sgra\ has changed on the time scales of years, while the first model fit assumes it remains constant over the considered time range. 

A change in flaring properties on years time scales is further supported by other lines of reasoning: the consistency of the light curves in 2010, 2011, and 2012 with a pure Poisson process; the non-zero probabilities for low $\zeta$ in the individual years making up set B, which cannot be fully explained by their low cadence alone (e.g. comparing 2019 with 2007 and 2008); and the shift in the probability density function of $\zeta$ between set $A$ and set $B$. We note that the best-fitted $\zeta$ values for those two sets do not change beyond the $1-\sigma$ confidence level. However, combined, the lines of evidence above are consistent with a decrease in flaring activity between 2008 and 2012. 

\begin{figure}
    \centering
    \includegraphics[width=\columnwidth]{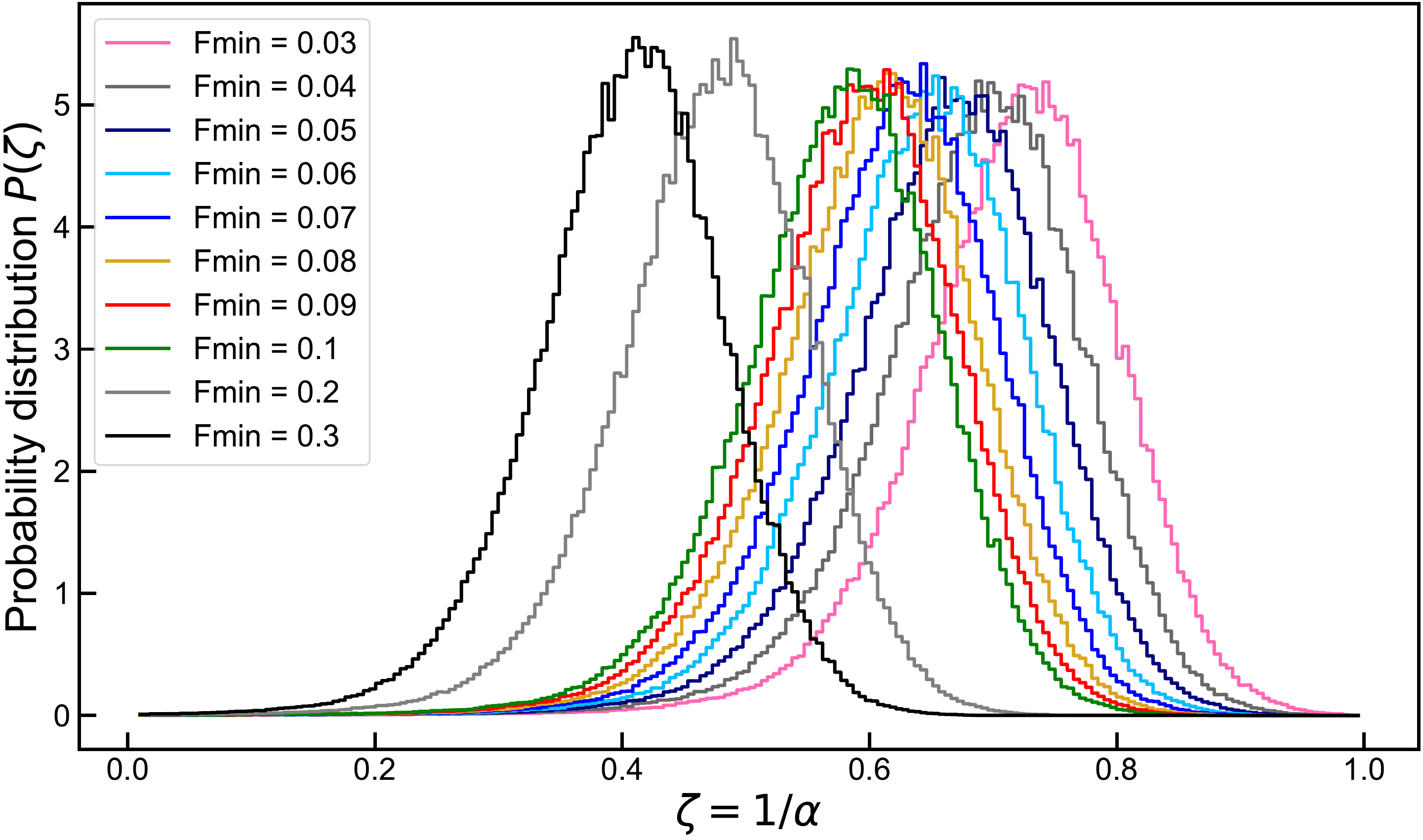} \\
    \includegraphics[width=\columnwidth]{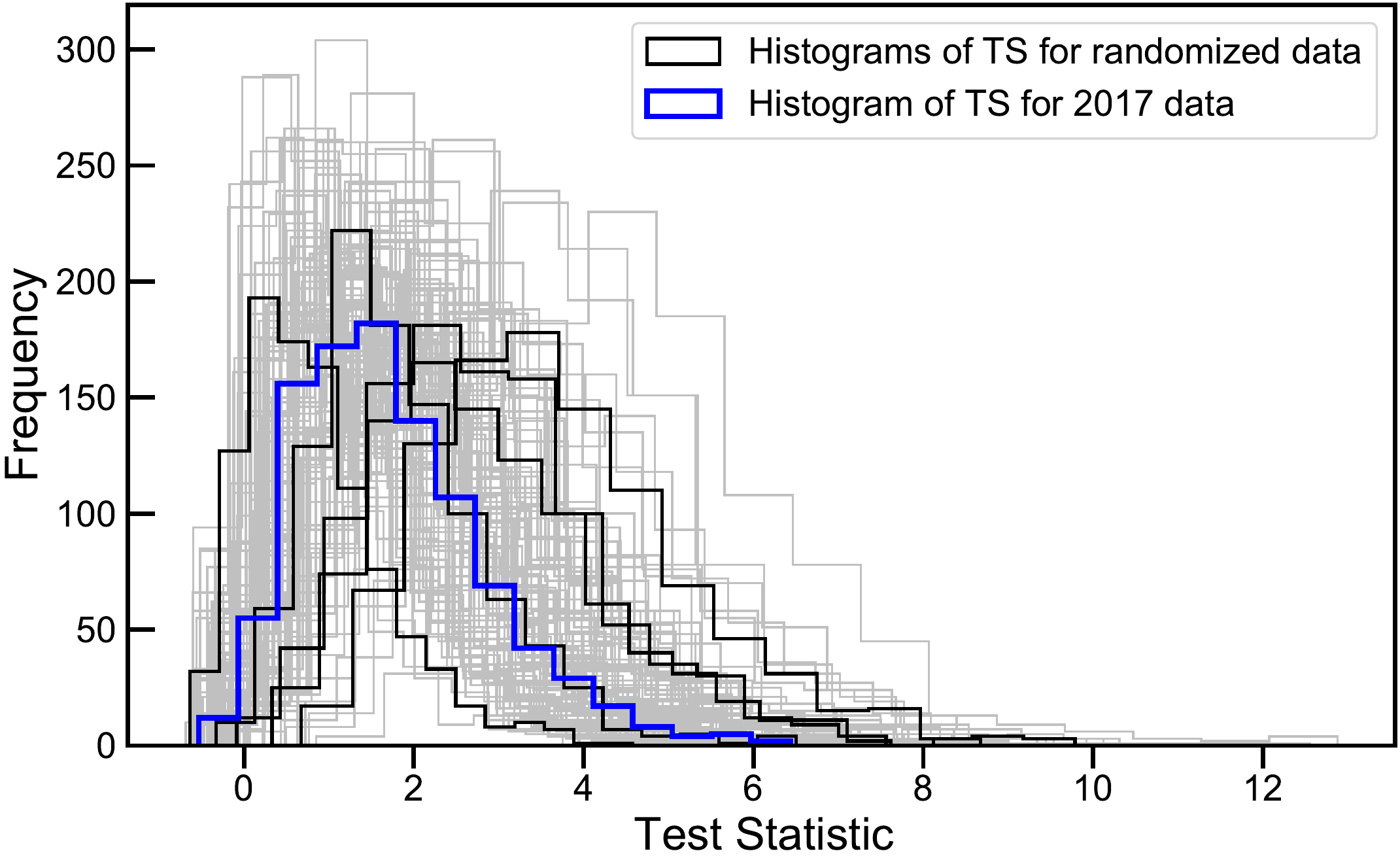}
    \caption{\textbf{Top panel:} Probability density as function of $\zeta$ using multiple values of $F_{\rm min}$ with 2017 data of \sgra. For a given $F_{\rm min}$, the best $\zeta$ corresponds to the maximum in the graph. \textbf{Bottom panel:} Distribution of TS values for the best fit $\zeta$ and $Q$ for 2017 data (blue) and for 100 randomized datasets with the same number of datapoints.}
    \label{fig:TSvsalpha}
\end{figure}

Before discussing such variations in the physical picture of Sgr A*, we briefly turn to the assumptions of the method. Firstly, the value of $F_{\rm min}$ also affects the fits and measured parameters in the data. Regarding the 1D method, in the top panel of Figure \ref{fig:TSvsalpha} we show again $p(\zeta)$ for the 2017 data, assuming multiple values of $F_{\rm min}$. The  $\zeta$ that best match the observations gives the maximum in each graph. It is noticeable that a higher $F_{\rm min}$ returns a lower $\zeta$. %and also a wider graph, indicating larger uncertainties when increasing $F_{\rm min}$. 
From Eq. \eqref{eq:1d} we see that as the value of $\zeta$ decreases, it reduces the value of $\overline{Q}$, and in fact, the 1D method breaks for any $F_{\rm min} > 0.30$, as this would require negative values of $\overline{Q}$ from Eq. \eqref{eq:1d}. The same effect, albeit at different exact values of $\zeta$ is observed in the analysis of any of the other years. Moreover, this also can be observed when the 2D method is applied. Since our study compares subsets of the \swift\ monitoring using the same $F_{\rm min}$, the dependence of $\zeta$ on $F_{\rm min}$ does not affect our qualitative conclusions. However, Figure \ref{fig:TSvsalpha} does show that the absolute value of $\zeta$ found in the analysis of a single dataset cannot be taken as a unique description of the underlying flare flux distribution. The best-fit values in Table \ref{tab:test} should therefore only be interpreted in relation to each other, preferably comparing the full distributions shown in Appendix \ref{appendixB}.

Not only $F_{\rm min}$ affects the measured values of $Q$ and $\zeta$; comparing the columns in Table \ref{tab:test} shows that the one- and two-dimensional approaches also find slightly different best fit parameters. The best-fit parameters are however consistent within their one-sigma uncertainties. Importantly, we find that the 1D-method is better able to constrain the $\zeta$-parameter. A possible reason for the difference in the exact best-fit value might be that the one-dimensional method ignores the uncertainty on the measured count rate: this uncertainty turns the line probed by the one-dimensional method in the $\zeta$-$Q$ plane, into a band, which is automatically fully explored in the two-dimensional method. 

We also notice that dividing the data into different sets, i.e. having smaller number of datapoints in each, makes the fits statistically better (i.e. reaching lower TS values). This happens also when we randomized the data for individual years, as described in Section \ref{sec:results} (testing the differences between sets A and B). We show this more explicitly in Figure \ref{fig:TSvsalpha} (bottom): in blue, we plot the histogram of TS values for the comparisons between the 2017 data and the $1000$ synthetic light curves generated using the best-fit values of $\zeta$ and $Q$ (see Table \ref{tab:test}). In the same figure, in grey, we plot the same histograms, for 100 light curves of the 2017 length, randomly picked from the full dataset; we also plot a small number these in black, to highlight their shape. The similarity in the distributions indicates that the statistical quality of the \textit{best} fits are similar, confirming that smaller light curves are easier to fit satisfactorily. 

However, we stress that the above analysis only compares the TS values for the best-fit parameters: whether changes between subsets of the data are found, depends also on whether a satisfactory fit is only found for a limited range of parameters. This comparison therefore does not imply that the fits are \textit{only} better due to the smaller number of data points, as discussed in Section \ref{sec:results} as well: even at low-cadence, synthetic light curves created using the best-fit parameters of 2006--2007 and 2017--2019 do not return $\zeta$ upper limits, and the comparison between 2008, 2009, and 2019, yields a similar suggestion. However, we cannot exclude the possibility that the lower cadence contributes to the better fits. 

Turning to the interpretation of the possibly variable flaring rate, we can first compare our work with the results from \cite{ponti2015}, which reports an increase in \sgra\ emission at the end of 2013 and 2014. Our work shows evidence that \sgra's emission in the X-ray band presents different properties between years, and although we are ignoring 2013--2016 data, the activity of \sgra\ in the last years of observation of \swift\ shows higher count rates than the rest of the years. On the other hand, work performed by \cite{yuan2015} and \cite{bouffard2019} reported no sign of changes in \sgra's activity using \chan\ observations during different epochs. Possibly the flaring rate of \sgra\ doesn't show measurable changes on time scales shorter than a year, as shown by the work from \cite{neilsen2015}. This would explain why such short-time-scale changes are not seen by either \swift\ or \chan,\ while they may occur on few-years time-scales. Hence we need multi-year, recurring monitoring to further test and confirm any of these results.

The nIR emission from Sgr A* also shows strong flaring variability, with fluxes that have been modelled as either a power law or log-normal distribution \citep{witzel2012,witzel2018} with an additional high-flux tail \citep{doddseden2011}, or a log-log normal distribution \citep{meyer2014}. Investigating the emission mechanism of the nIR and X-ray flares, \citet{neilsen2015} assume that the X-ray and nIR fluxes both follow power law distributions and that these fluxes are coupled. These simple assumptions imply a direct relation between the power law indices in nIR and X-rays. In that scenario, the observed variability in X-ray flaring rate would be accompanied with similar changes in the nIR flaring rate. However, if such changes in the nIR are not present, the nIR -- X-ray coupling either changes over time, or the X-ray and nIR flares are not (always) related. If the X-ray and nIR flares are indeed manifestations of the same underlying emission process, we deem it unlikely that the coupling between wavelengths varies over years time scales and instead expect a similar increase in nIR flaring rate as the X-ray activity increases.

The nIR flux distribution of Sgr A* indeed varies on time scales of years; for instance, \citet{doddseden2011} and \citet{witzel2012} both analyse overlapping, but not identical, sets of \textit{VLT/NACO} observations and report different flare flux distributions (both in shape and parameters). More recently, analysing \keck\ nIR observations obtained in April and May 2019, \citet{do2019} reported an increase in flaring rate compared to \keck\ observations between 2005 and 2013. The increased 2019 flaring rate coincides with the increased flaring rate seen in our \swift\ monitoring, while during the 2005--2013 interval, the \swift\ monitoring is dominated by low flaring rate years (2008--2012, compared to 2006--2007 with a higher flaring rate). These comparisons are suggestive of a coupled change in flaring rate between nIR and X-rays, which could be confirmed by a time-resolved analysis of the nIR and X-ray fluxes on year time scales in future data. 

Semi-analytical studies generally favour synchrotron emission from a non-thermal population of electrons as the source of simultaneous nIR/X-ray flaring \citep[e.g.,][]{Dibi_2016}. The cooling of synchrotron electrons in these models results in the steepening of the spectral slope from the nIR to the X-rays and therefore, provides a crucial link between the two flare populations \citep[e.g.,][]{doddseden2010,Dibi_2014,ponti2017}. Magnetised blobs of plasma that naturally form in the turbulent accretion flow \citep[e.g.,][]{sironi2015,Ripperda2021} are said to be a viable source of rapid flaring activity in Sgr A* \citep[e.g.,][]{Ball_2016,Chatterjee2021,Scepi2021}. Indeed, \citet{Gutierrez_2020} suggests that a large enough magnetised blob could produce the high levels of non-thermal synchrotron emission that would be required to explain the \citet{do2019} nIR flare\footnote{We note that the flare visible in May 2019 in Swift monitoring (Figure \ref{fig_lc}) did not occur simultaneously with the \citet{do2019} nIR flare.}. Furthermore, an increase in the accretion rate could result in a strongly magnetised accretion disk as more and more magnetic fields are advected in. Such a disk is able to produce magnetised blobs with relativistic temperatures \citep{Ripperda_2020,Ripperda2021} that leads to stronger flares. Indeed, an increase in the X-ray flare rate in the period of 2017-2019 does seem to support this possibility. Current numerical models of accreting black holes are only able to address a few 10s of hours of Sgr A* activity \citep[e.g.,][]{Chan_2015,Ball_2016,chael2018,Chatterjee_2020}, and much longer simulations are necessary to predict whether a small change in the amount of accreted material on the timescale of years could trigger a large non-thermal event close to the black hole.

\citet{do2019} also suggest an alternative explanation for the observed increase in nIR activity: they suggest it could be related to the periastron passage in May 2018 of the windy star S0-2. However, if the increase in nIR and X-ray activity are indeed coupled, this explanation is inconsistent with the increased X-ray activity already observed in 2017 \citep[see also][]{ressler2018}. Alternatively, the increase in activity might result from the periastron passage of the gaseous object G2 in 2014 \citep{gillessen2012,gillessen2013a,gillessen2013b,madigan2016}. Simulations of the response of Sgr A* to this periastron passage predicts a delay in increased activity of a few to $\sim 10$ years \citep{schartmann2012, kawashima2017}. Without information from \swift\ between 2012 and 2017, we can infer an upper limit on the possible X-ray response to the G2 passage of $\sim 3$ years, lasting for $2$ years at the time of writing, thus fitting with those model predictions. 

\swift\ monitoring also suggests a higher flaring rate before 2008, followed by a low activity period between 2008 and 2012. Therefore, if the G2 object is responsible for the current high levels of activity, a similar object could have passed close by Sgr A* during the late 1990s or early 2000s to explain earlier enhanced activity. Indeed, a gaseous object similar to G2 \citep{clenet2004a,clenet2004b,clenet2005,ghez2005}, named G1 by \citet{pfuhl2015}, passed similarly close to Sgr A* in 2001 \citep{witzel2017}. We stress that these considerations are highly speculative; for instance, the \swift\ monitoring is only consistent with an increased flaring rate, but not with a change in the steady accretion activity of Sgr A*, although this might be hidden in the diffuse X-ray emission unassociated with Sgr A* due to \swift's point spread function. Also, \citet{yuan2015} did not find evidence for changes in X-ray activity observed by \textit{Chandra} after the passage of G2. However, if the enhanced activity is related to passages of objects such as G1 and G2, the decrease of flaring activity in 2008 -- roughly $7$ years after G1's periastron passage -- would predict an end to the current high level of flaring activity around $\sim 2021$. 

With continued monitoring in X-rays (by \swift) and nIR, the above prediction could be tested, although we note the importance of a regular and high cadence of these observations, similar to the 2017--2019 data, in order to eliminate any possible effects of sample size. Alternatively, new \textit{Chandra} observations could similarly help to assess the variability on years time scales, allowing for a comparison with the 2012 campaign. Finally, we remind the reader how this work focused on the broad statistical properties of the data set, and did not analyse individual flares; detailed statistical flare searches in the long-term \textit{Swift} data, such as performed by \citet{ponti2015}, could be an alternative route to explore this Sgr A* data set.

\section{Conclusions}
We have analysed the empirical CDF of the X-ray emission from \sgra\ using \swift\ data for 2006--2012 and 2017--2019. By assuming the X-ray emission of \sgra\ is composed of a quiescent and a flaring component, we modelled it as the sum of a constant flux and a power law flux distribution, respectively. We adopted two different methods to fit the CDF, yielding consistent results: a 2D method allowing the flare parameter $\zeta$ and the quiescent parameter $Q$ to vary independently, and a 1D method establishing a dependence between $\zeta$ and $Q$ as shown in Eq. \eqref{eq:1d}. We found that the full dataset cannot be correctly described with this model, while individual years do follow it better.

More concretely, we found that the data can be divided in two different groups: (1) a set of data with high flaring rate (2006--2007 and 2017--2019), which we called set $A$, and (2) a set of data with low flaring rate (2008--2012), labelled as set $B$. The main difference between these to sets can be summarized as: 

\begin{enumerate}
    \item All years forming set $A$ have a well constrained value of $\zeta$, as shown in bottom panel of Figure \ref{fig:contour_plots}. Therefore, the model of constant $+$ power law distribution can describe the data of these years. However, the years of set $B$ present larger errors on $\zeta$ or it cannot be constrained and only an upper limit can be established. In some cases in set $B$, the emission can be well described by a pure constant model with no flaring at all.  
    
    \item The values of $\zeta$ for set $B$ may be systematically shifted to lower values than the ones obtained for set $A$. Such lower values of $\zeta$ would indicate a lower flaring rate for set $B$. While we highlight that these changes are within the 1-$\sigma$ confidence level, they may hint towards a change in the flaring rate between different years.
    
\end{enumerate}

Finally, we tested explicitly the effect of dividing the data into two or more sets and we found that the probability of the improvement in the fits is due by chance when reducing the number of data points is just $\sim 0.3\%$. Similarly, the poorly-constrained parameters for some the years in set $B$ cannot be explained by only the lower number of data points of those sets. Therefore, we interpret these results as evidence for a change in the flaring rate of \sgra, being more active during 2006--2007 and 2017--2019, compared with the years of 2008--2012.

\section*{Acknowledgements}
The authors thank the referee for an insightful and constructive report that improved the quality of this work. AA, JvdE and ND were supported via an NWO/Vidi grant awarded to ND. AA gratefully acknowledges support from the ASPIRE program and thanks the University of Amsterdam for hospitality. JvdE is also supported by a Lee Hysan Junior Research Fellowship from St Hilda’s College, Oxford. KC and SM are supported by the NWO/VICI grant (no. 639.043.513) awarded to SM. KC is additionally supported by a Black Hole Initiative Research Fellowship from Harvard University, funded by the Gordon and Betty More Foundation, John Templeton Foundation and the Black Hole PIRE program. GP acknowledges funding from the European Research Council (ERC) under the European Union’s Horizon 2020 research and innovation programme (grant agreement No 865637). COH is supported by NSERC grant RGPIN-2016-04602. DA acknowledges support from the Royal Society. This work made use of public data from the \swift\ data archive, and data supplied by the UK \swift\ Science Data Center at the University of Leicester. 

\section*{Data availability}

A data reproduction package for this paper can be found at the following DOI: 10.5281/zenodo.5140005

\footnotesize{
%\bibliographystyle{mn2e}
%\bibliography{sgra}

}

\appendix 

\section{Anderson-Darling test statistics and significance levels}
\label{appendixA}

\begin{figure}
    \centering
    \includegraphics[width=\columnwidth]{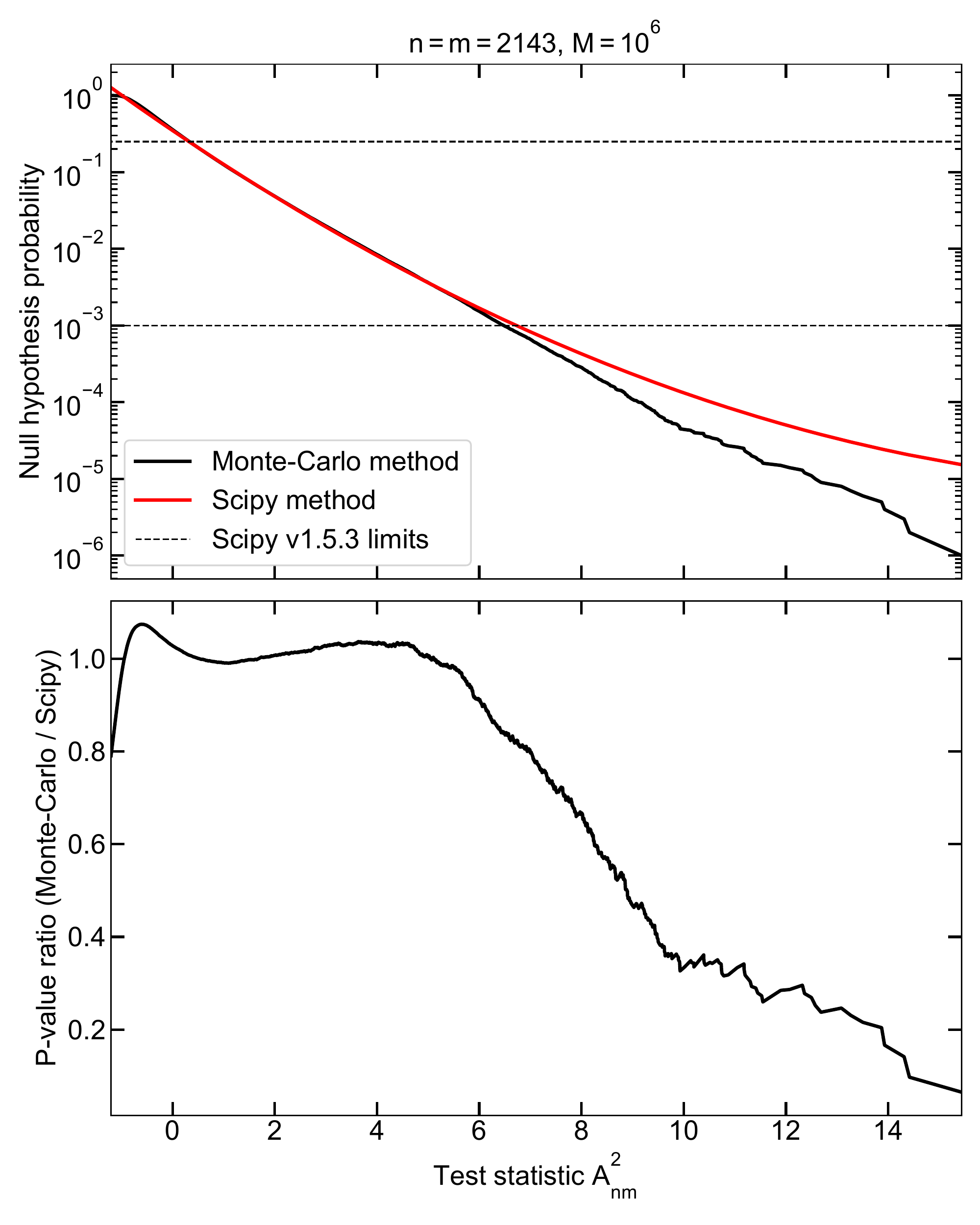} \\
    \caption{\textbf{Top panel:} Null hypothesis robability as function of Anderson-Darling test statistic, using both \textsc{scipy.stats.anderson\_ksamp} v1.1.0 (red) and a Monte-Carlo method (black). The dashed lines indicate the limiting p-values in \textsc{scipy} v.1.5.3. \textbf{Bottom panel:} the ratio between the probabilities calculated via the two methods plotted in the upper panel.}
    \label{fig:app}
\end{figure}

In order to test whether the generated synthetic light curve for a given parameter pair ($\zeta$, $Q$) and the observed \textit{Swift}/XRT light curve are consistent with the same underlying flux distribution, we employ the Anderson-Darling (AD) test \citep{anderson1954}. As noted by \citet{neilsen2015} in their work on the \textit{Chandra} XVP data, and in earlier work by \citet{scholz87}, this test is more sensitive for deviations in the tails of the distributions than for instance the Kolmogorov-Smirnov test, making it more suitable for comparing flaring behaviour on top of a second, quiescent component. However, a downside to the use of the AD test is the lack of a single analytic expression relating the observed test statistic, referred to as TS in the main paper and $A^2$ in statistical literature, to a p-value: the probability that the two populations arise from the same underlying distribution. 

Instead, early studies of the AD-test and its k-sample extensions calculated critical values for different sizes of the compared samples. In the 2-sample case, these critical values are defined as the test statistic $A^2_{\rm crit}$ where $p(A^2_{nm} \geq A^2_{\rm crit}) < \alpha_{\rm crit}$, where $\alpha$ is the confidence level \citep[we use the typical notation in statistical literature, not to be confused with the flaring parameter from][]{neilsen2015}, and $n$ and $m$ are the sizes of the two compared samples. The critical values change as a function of the sample sizes and are often calculated for $\alpha_{\rm crit} = 0.01, 0.05, 0.10$ \citep{pettitt76, scholz87}. Analytic extrapolations of these critical values can then be used to either find critical values for larger sample sizes $n$ and $m$, or convert test statistics to probabilities for non-critical values. The \textsc{python} \textsc{scipy} function \textsc{scipy.stats.anderson\_ksamp}, which we used to calculate the AD test statistic throughout this work, uses such extrapolations to return probabilities as well. However, these probabilities are capped at $p=0.001$ and $p=0.25$, limiting the use of this function in calculating AD probabilities. 

An alternative route to convert the test statistics to probabilities is via Monte-Carlo calculations. For two samples of sizes $n$ and $m$, one can explicitly calculate the distribution of $A^2_{nm}$ by calculating $A^2_{nm}$ for all $(n+m)! / (n!m!)$ possible orders of all $(n+m)$ values. The probability associated with the measured test statistic $A^2_{\rm measured}$, i.e. $p(A^2_{nm} \geq A^2_{\rm measured})$, can then be calculated from the distribution. This approach has a significant limitation: the number of orderings $(n+m)! / (n!m!)$ quickly increases, yielding unfeasible computational times for our data sets: our sample sizes exceed $100$ for all years except 2009, implying both $n$ and $m$ exceed $100$, while $n=m=100$ yields $\sim 10^{59}$ orderings. However, as \citet{scholz87} note, one can instead generate a large number of randomly picked orderings, and use their $A^2_{nm}$-values as an approximation of the full underlying distribution -- the probabilities calculated from this Monte-Carlo approach are an unbiased estimator of the real probability. 

We therefore applied this Monte-Carlo approach to convert measured test statistics to p-values. As the AD-test is a rank test, $A^2_{nm}$ only depends on the order of the values of the two samples, and not on the values themselves. The null hypothesis in our analysis is that both samples are drawn from the same underlying distribution. Therefore, since $A^2_{nm}$ does not depend on the values, we can define a list of integers ranging from $1$ to $n+m$ as the underlying sample. In our analysis, $n$ and $m$ are equal, as we are comparing an observed light curve (for one year, set A/B, or the full data set) with a synthetic one of equal length. For each Monte-Carlo iteration, we then create two samples, each of size $n$ (=$m$=half the full integer list), by randomly choosing integers from the list. We then calculate and save their $A^2_{nm}$ using \textsc{scipy.stats.anderson\_ksamp} and repeat $M$ times, where $M$ is set by the intended accuracy in p-values. 

To show the importance of using Monte-Carlo calculations, we show a comparison between the test statistic to p-value conversion using \textsc{scipy.stats.anderson\_ksamp} and our approach in Figure \ref{fig:app}. We show the results for $M=10^6$ iterations using $n=m=2143$, which corresponds to the complete data set. The dashed lines indicate the minimum and maximum returned p-value using the most recent \textsc{scipy} (v1.5.3), while we used an older version (v1.1.0) without this restriction for comparison to show the effects at smaller and larger test statistics. Between the two limits, both approaches agree reasonably well, although the effect of applying a analytic interpolation is visible in the variable ratio in the lower panel. The discrepancy grows especially above test statistics of $\sim 5$. As our method requires an accurate measurement of the p-values even for poorly-fitting parameters, in order to sample the entire parameter space, applying the Monte-Carlo simulations is essential.

\section{Parameters estimation and CDFs for individual years}
\label{appendixB}

\begin{figure*}
    \centering
    \includegraphics[width=0.66\columnwidth]{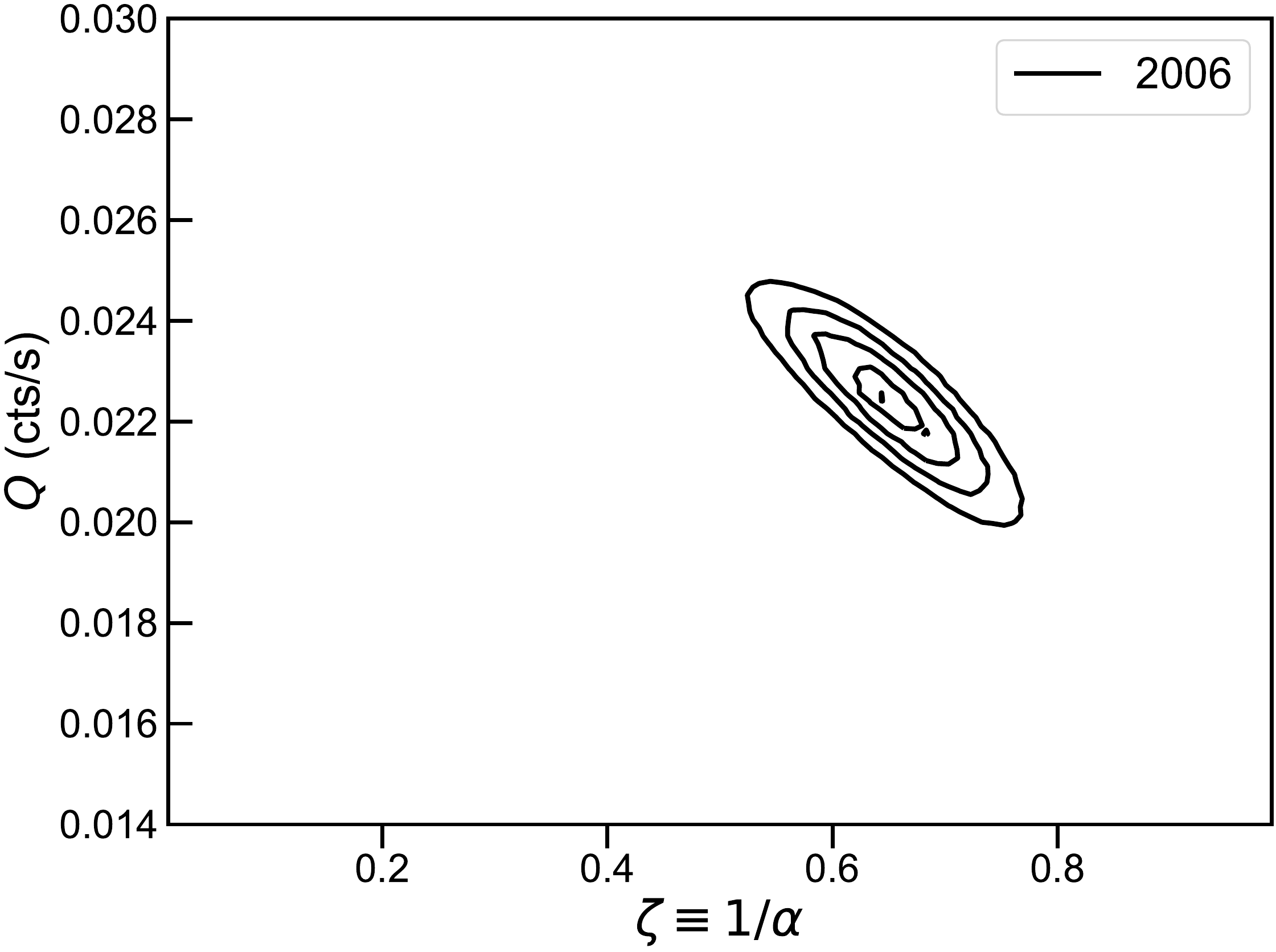}
    \includegraphics[width=0.66\columnwidth]{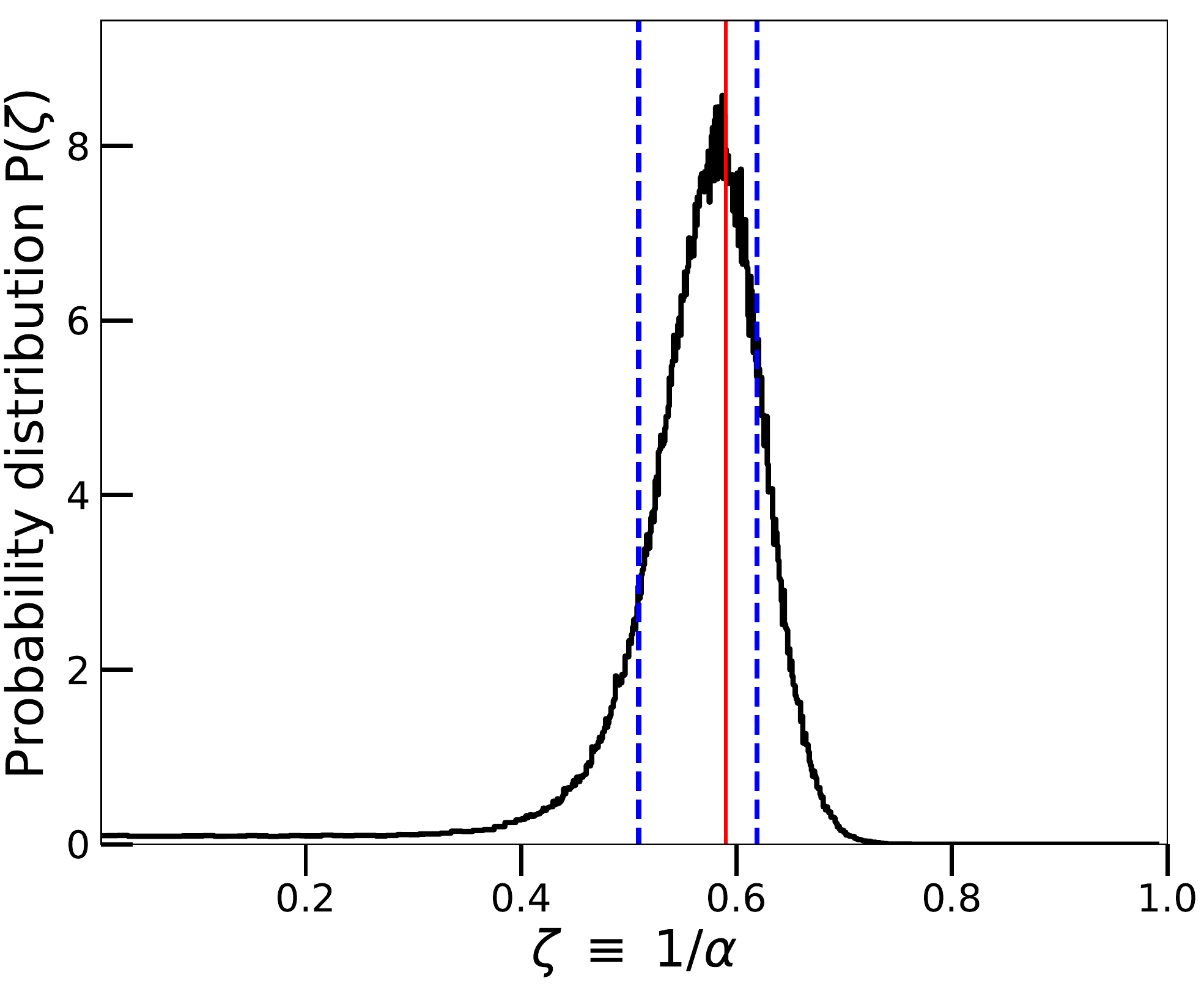}
    \includegraphics[width=0.72\columnwidth]{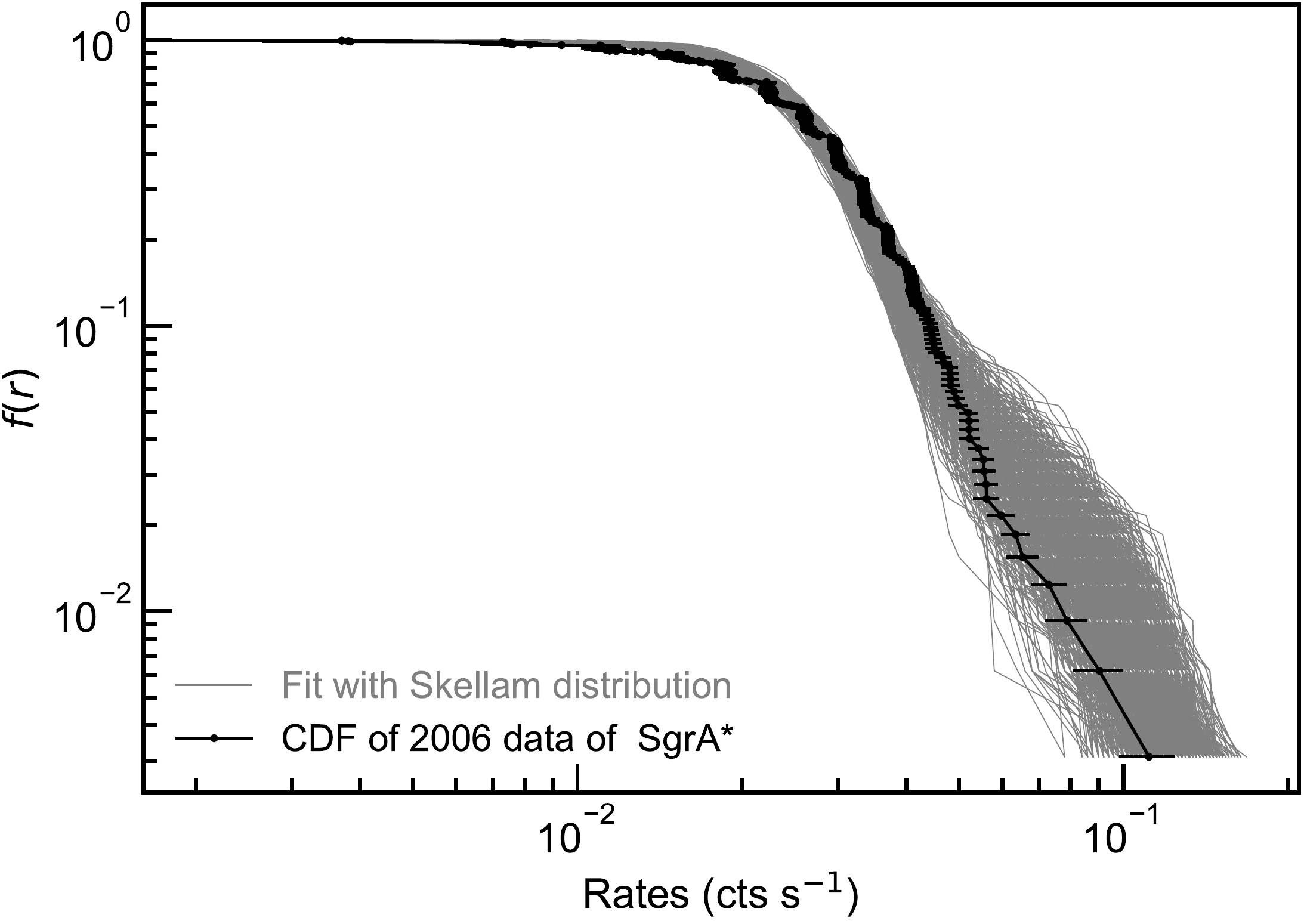} 
    
    \includegraphics[width=0.66\columnwidth]{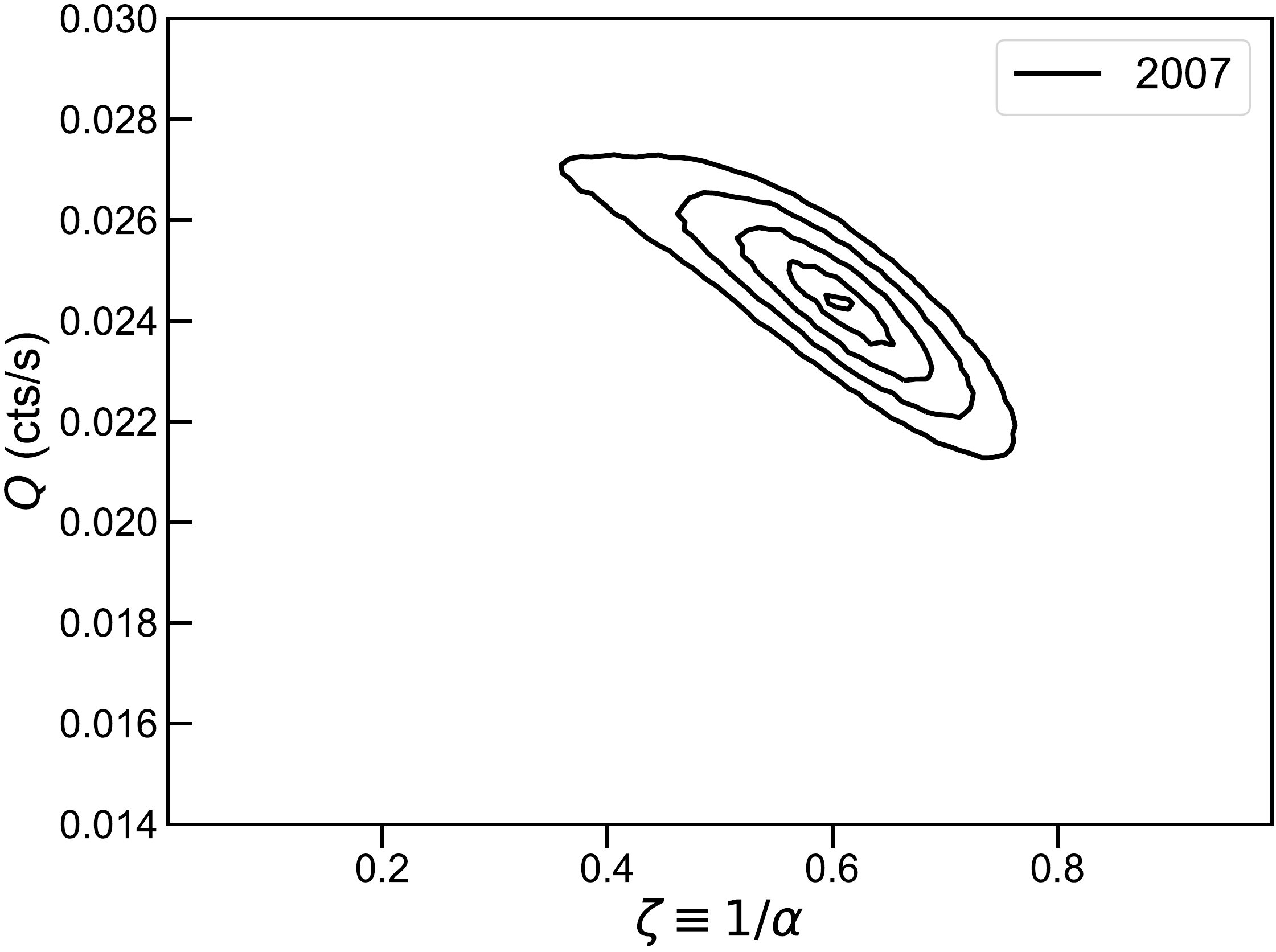}
    \includegraphics[width=0.66\columnwidth]{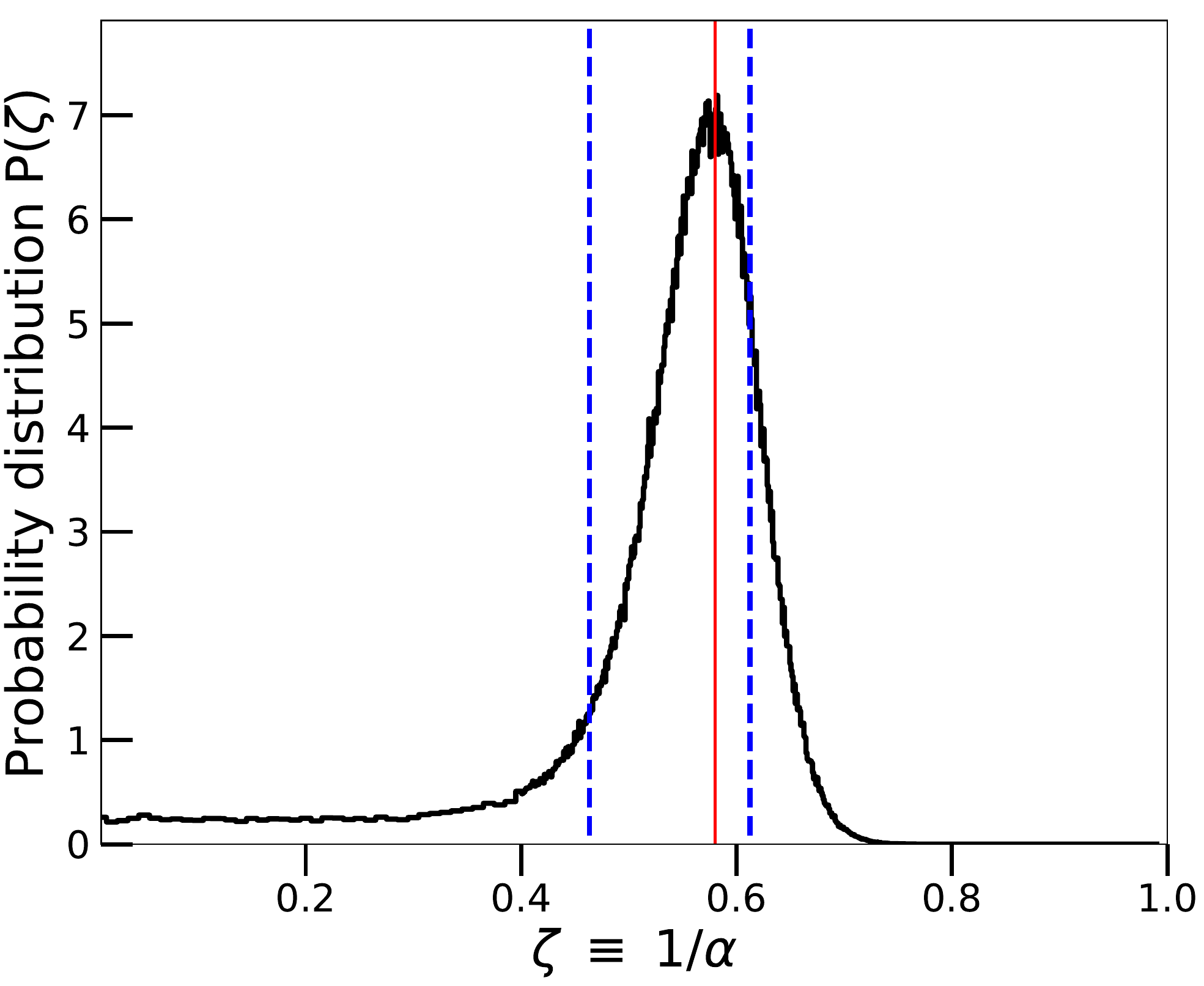}
    \includegraphics[width=0.72\columnwidth]{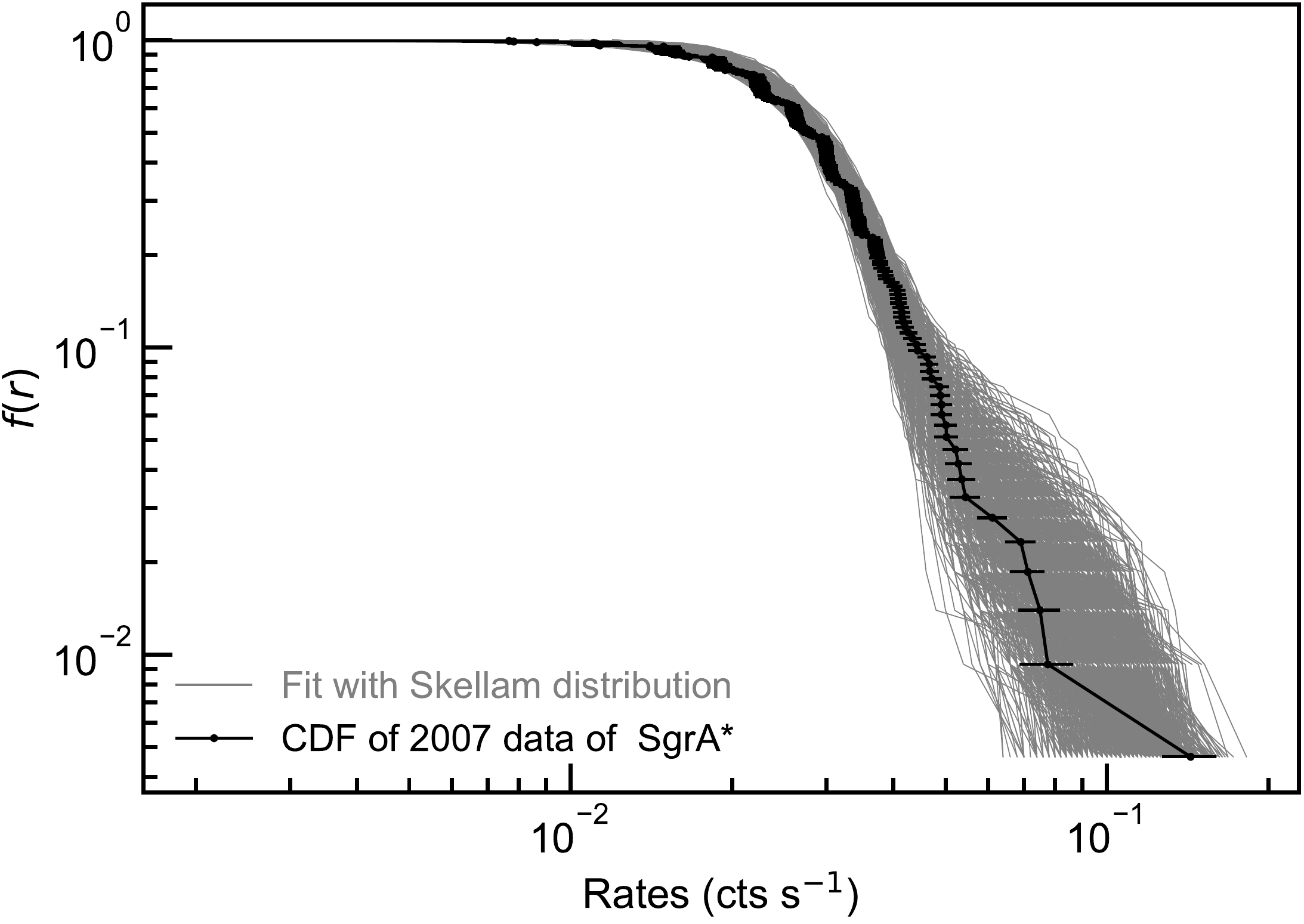} 
    
    \includegraphics[width=0.66\columnwidth]{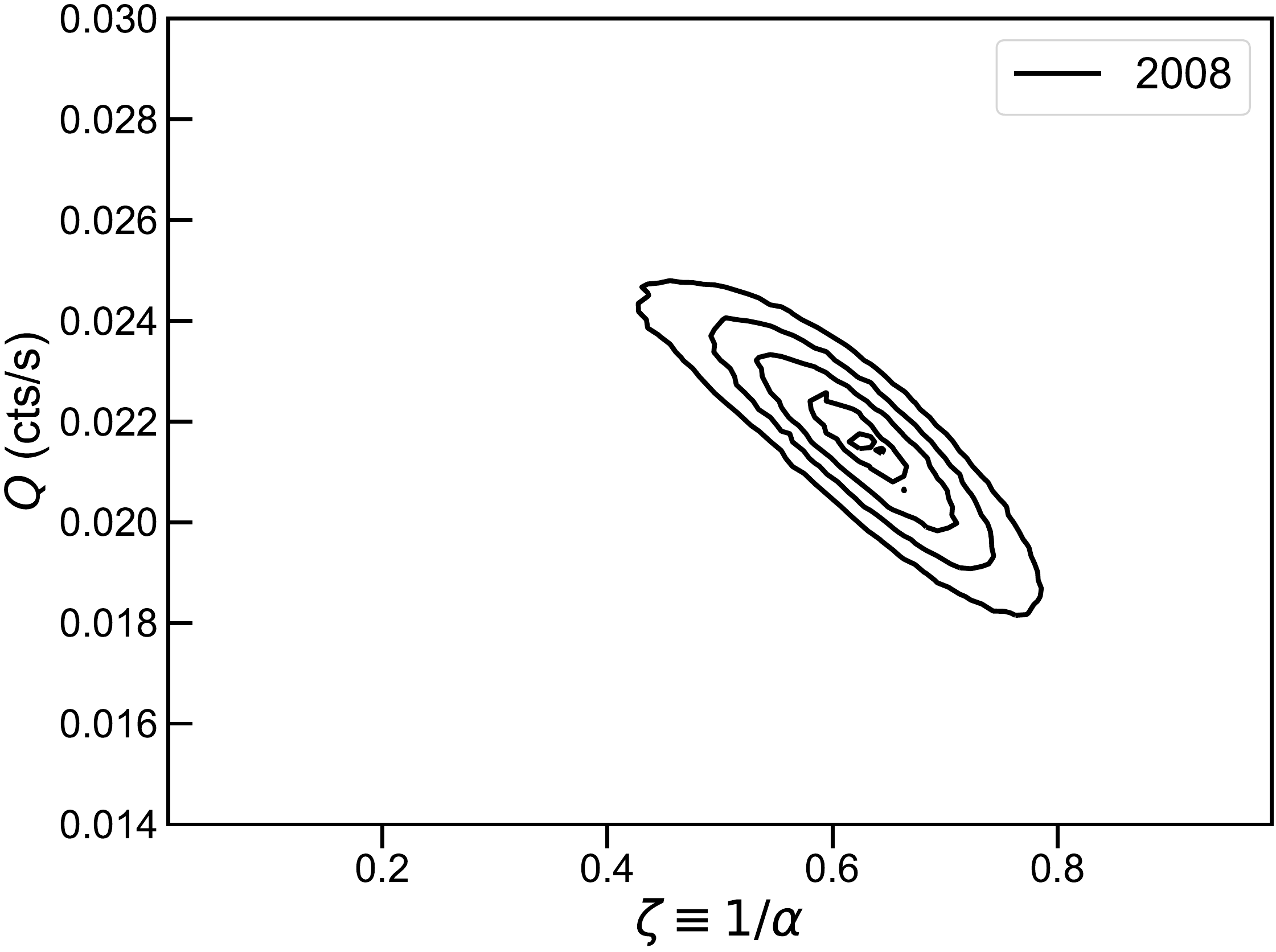}
    \includegraphics[width=0.66\columnwidth]{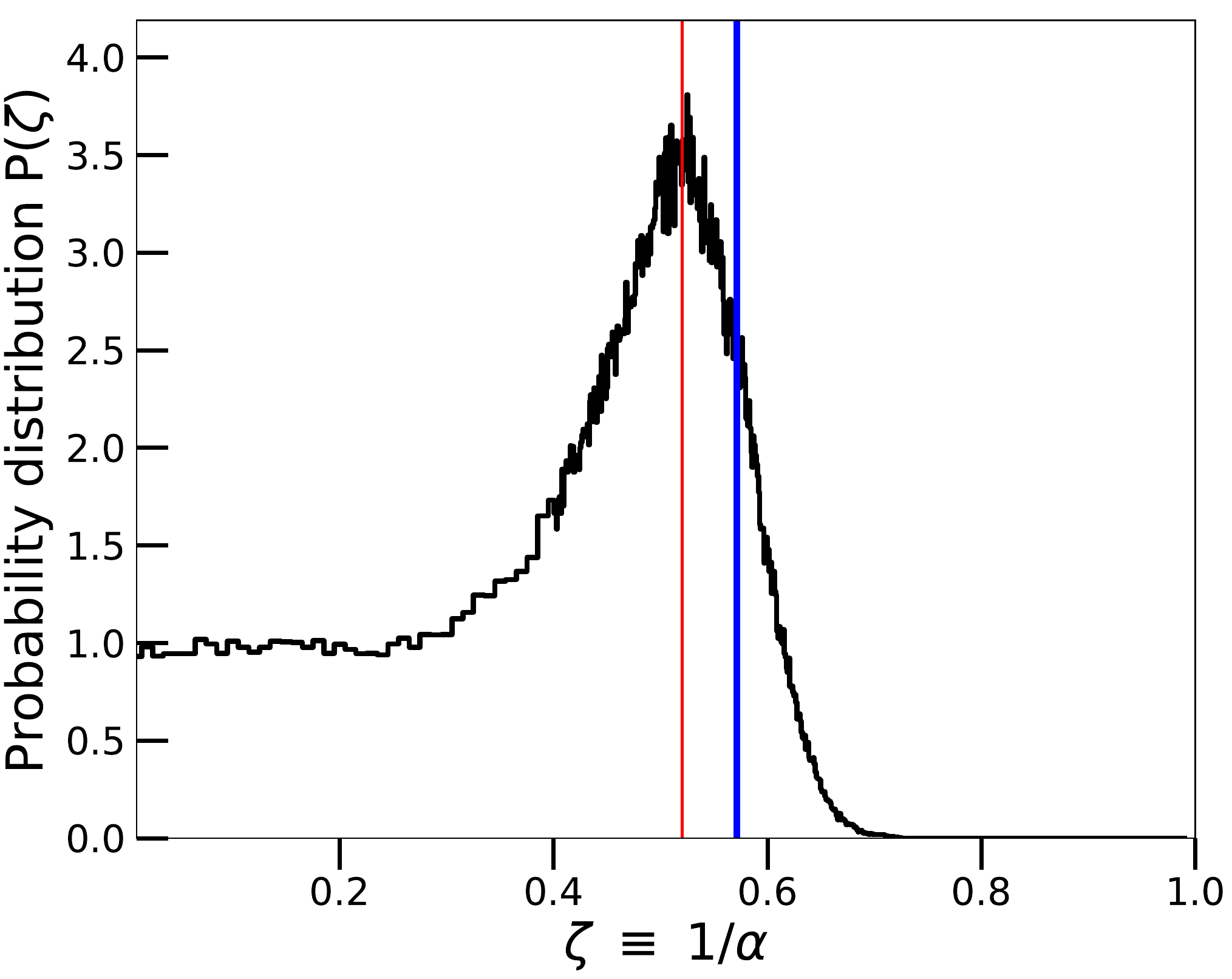}
    \includegraphics[width=0.72\columnwidth]{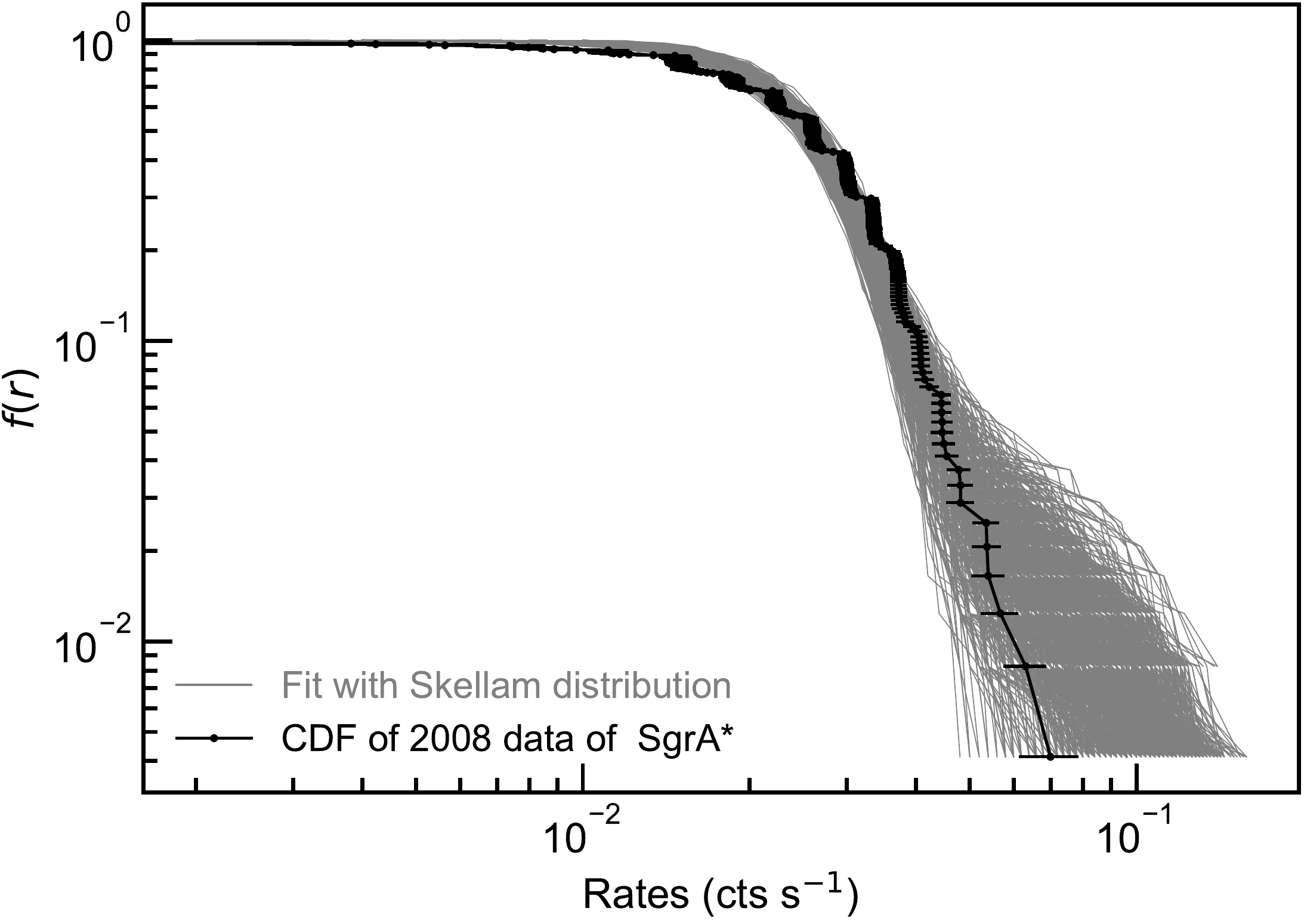} 
    
    \includegraphics[width=0.66\columnwidth]{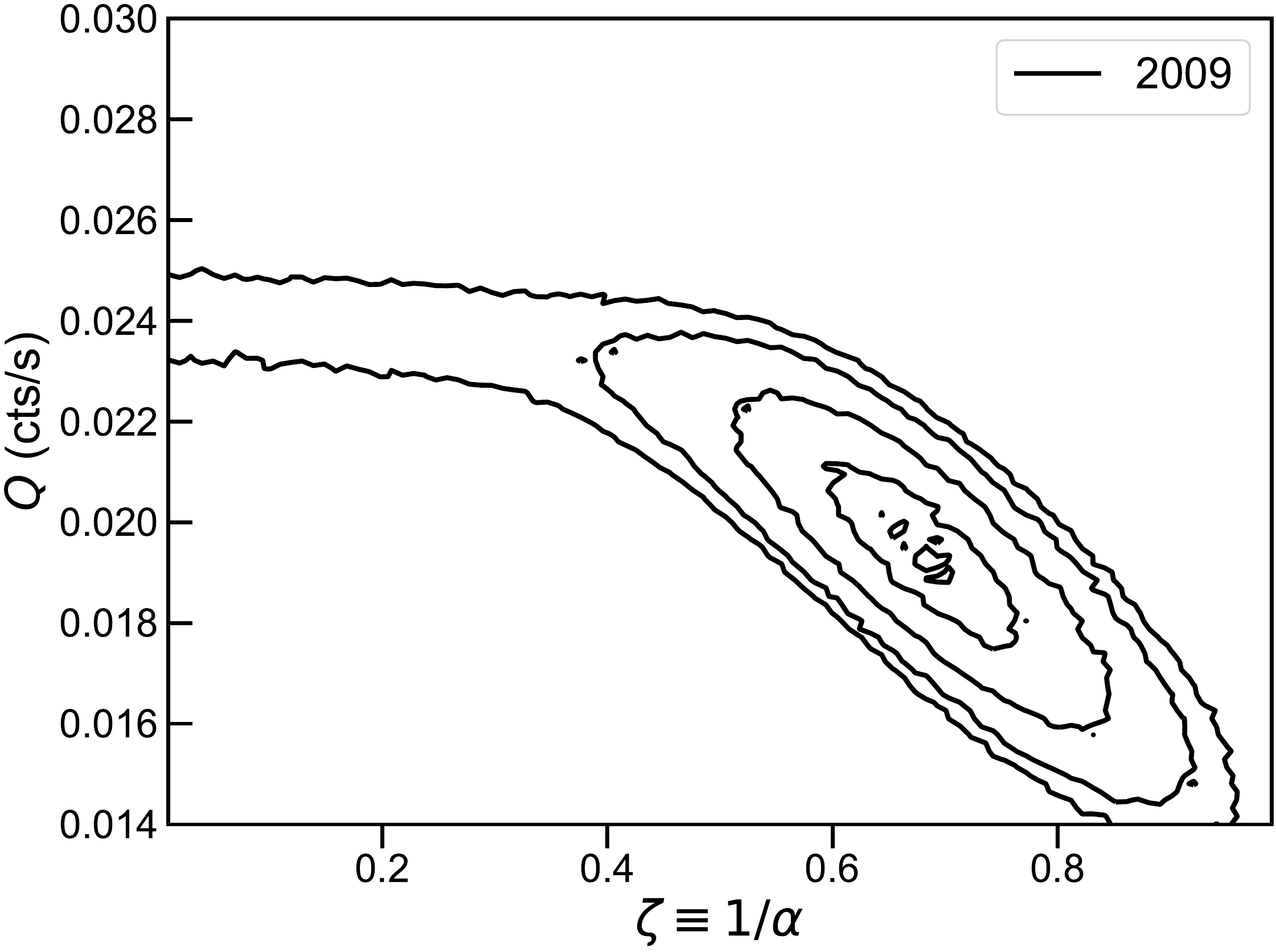}
    \includegraphics[width=0.66\columnwidth]{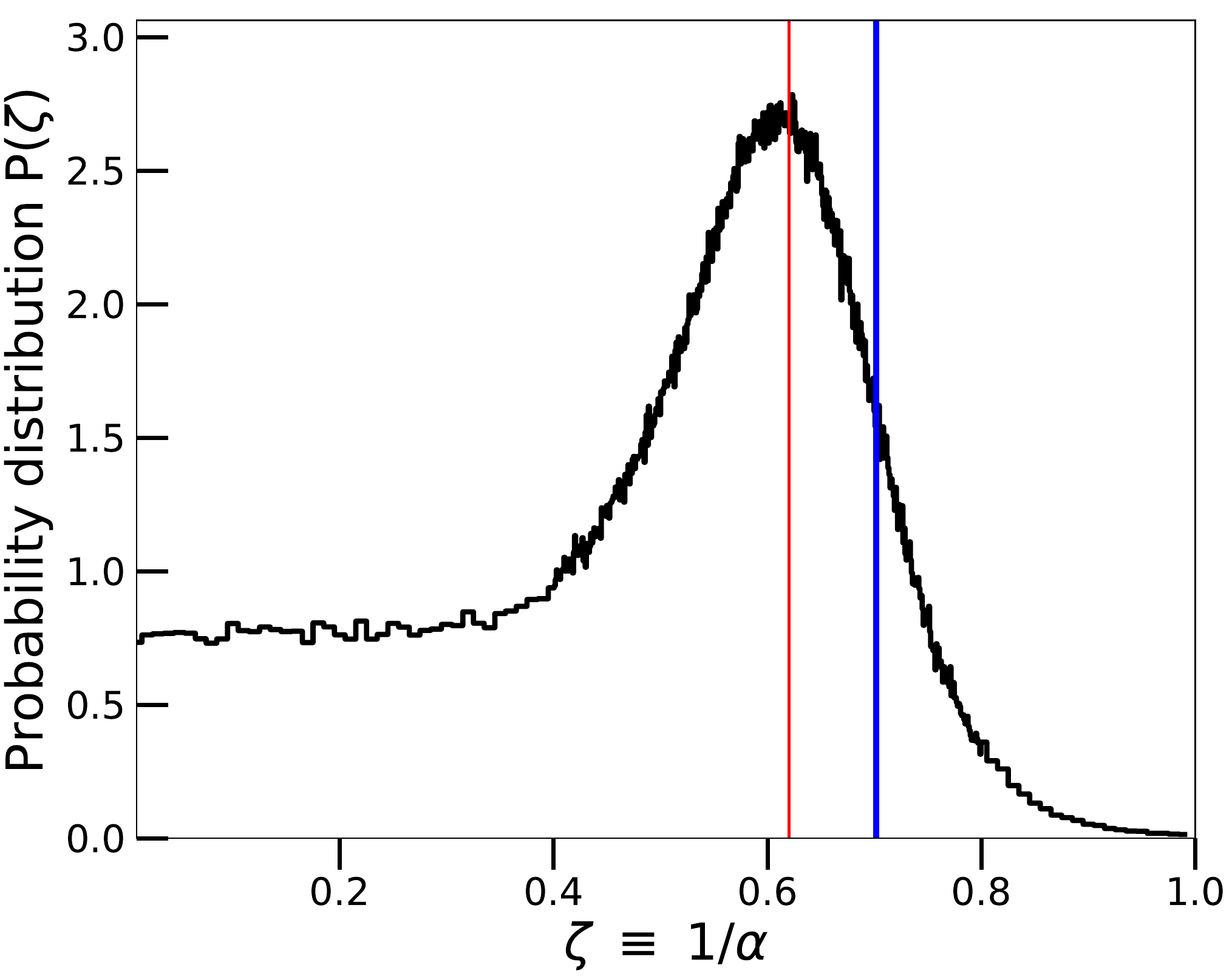}
    \includegraphics[width=0.72\columnwidth]{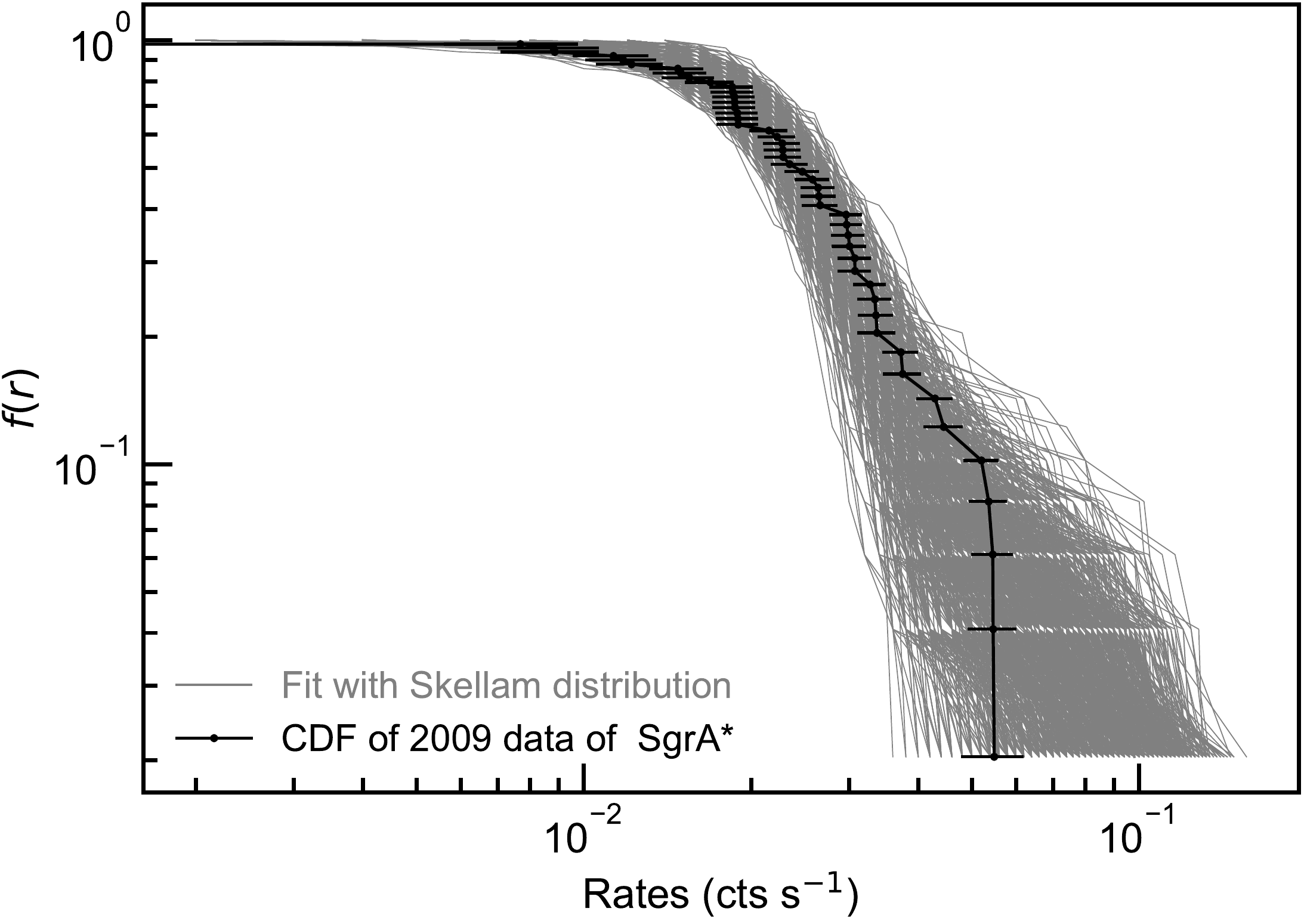} 
    
    \includegraphics[width=0.66\columnwidth]{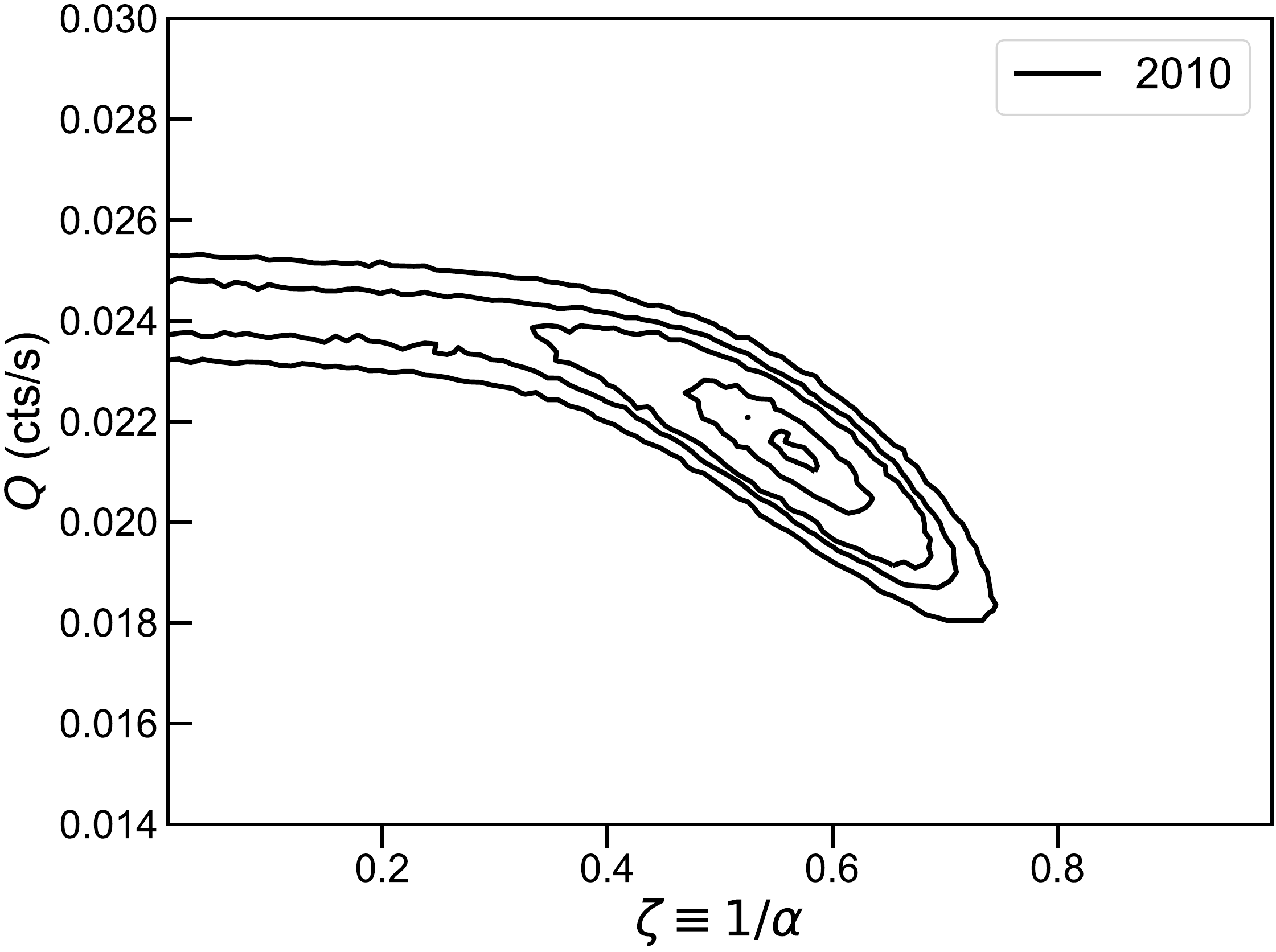}
    \includegraphics[width=0.66\columnwidth]{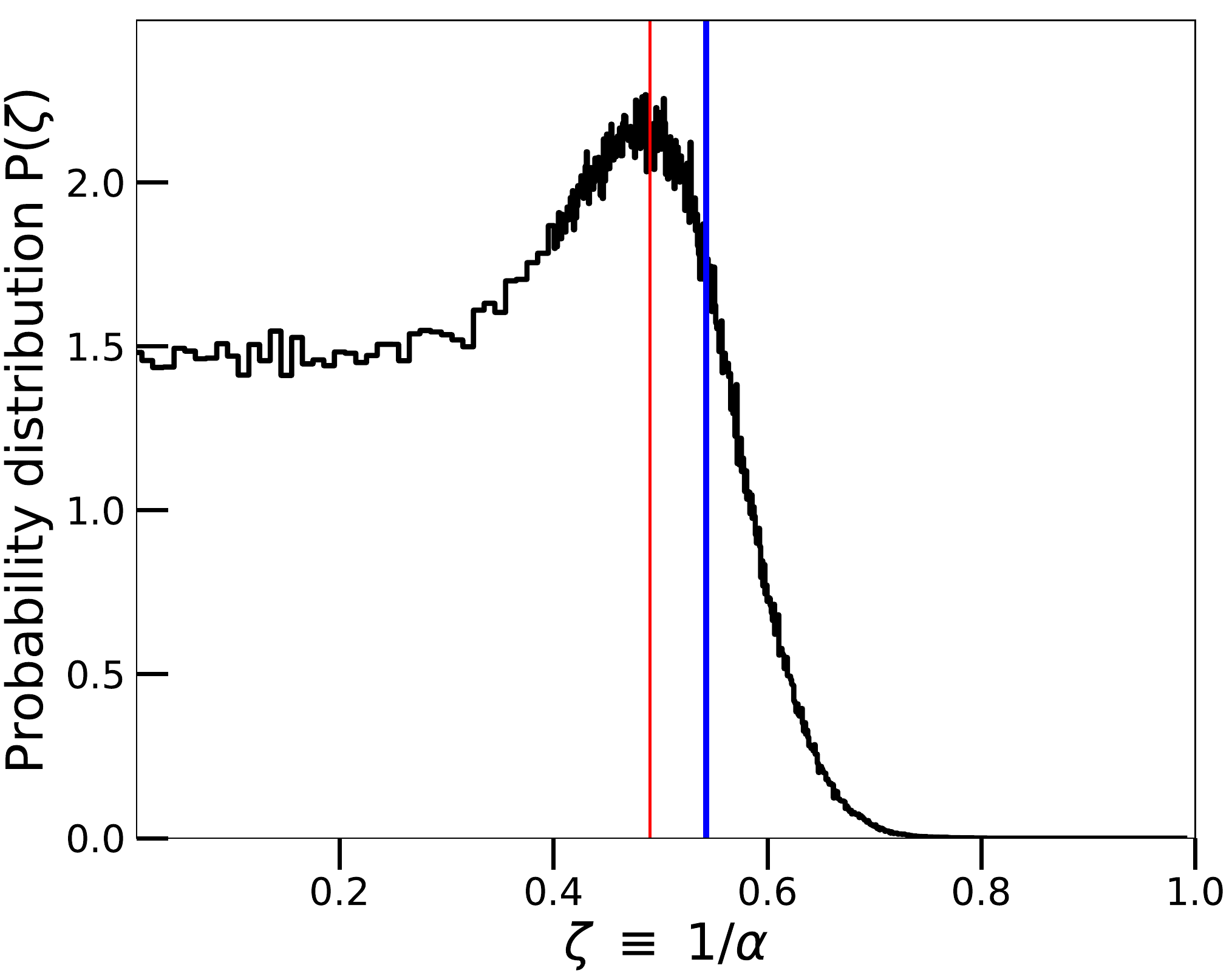}
    \includegraphics[width=0.72\columnwidth]{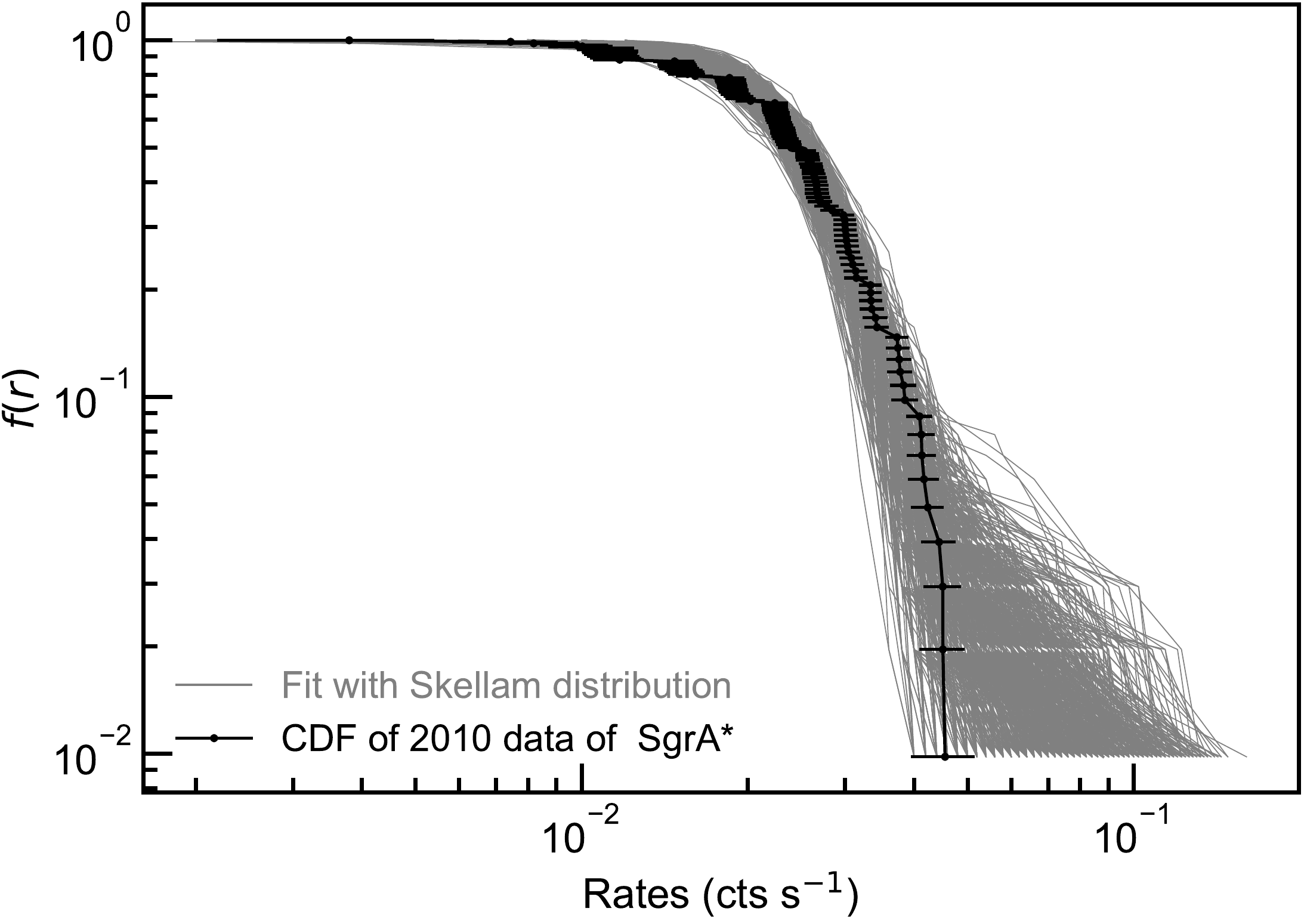} 

\end{figure*}

\begin{figure*}
    \centering
    \includegraphics[width=0.66\columnwidth]{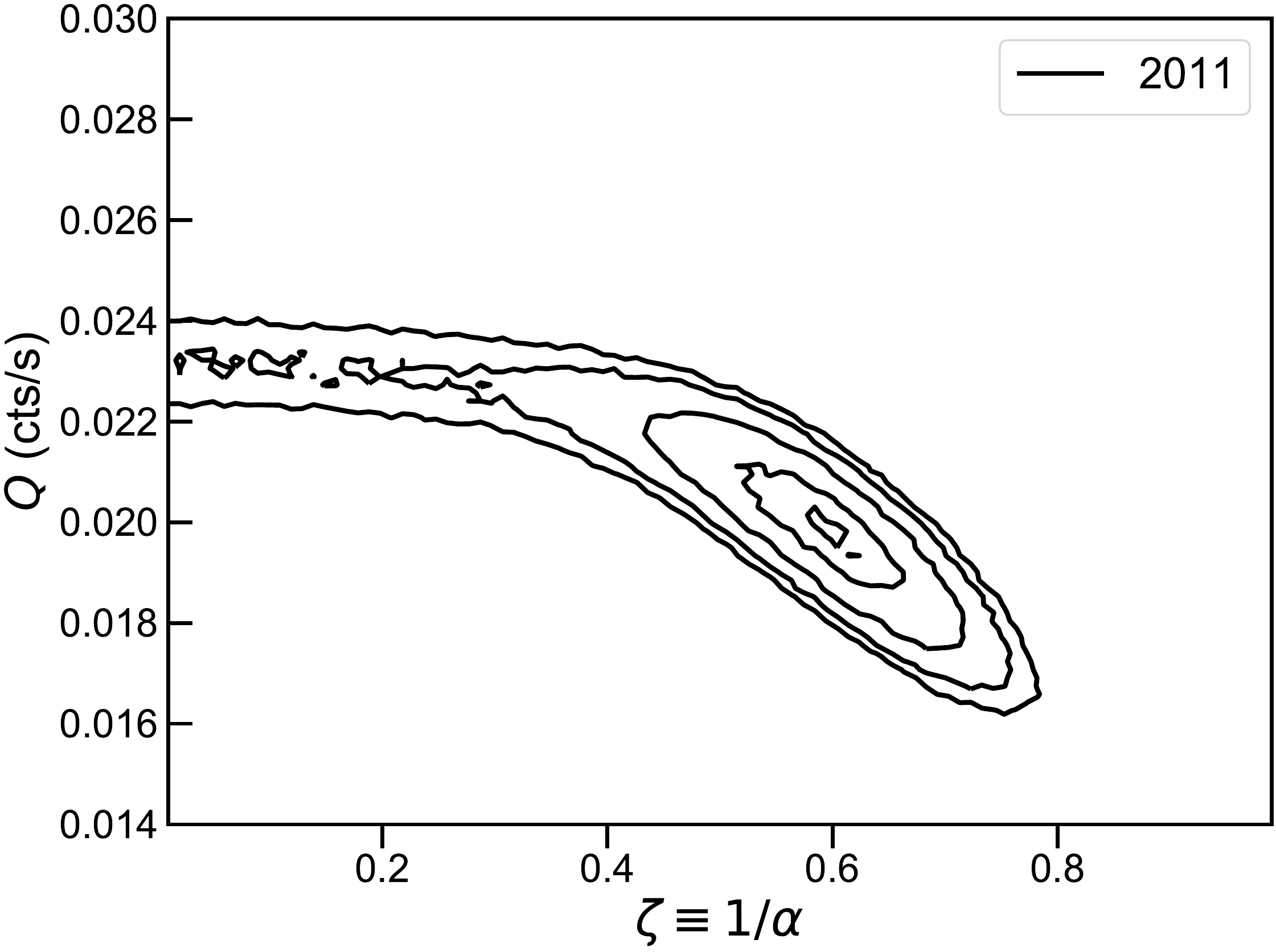}
    \includegraphics[width=0.66\columnwidth]{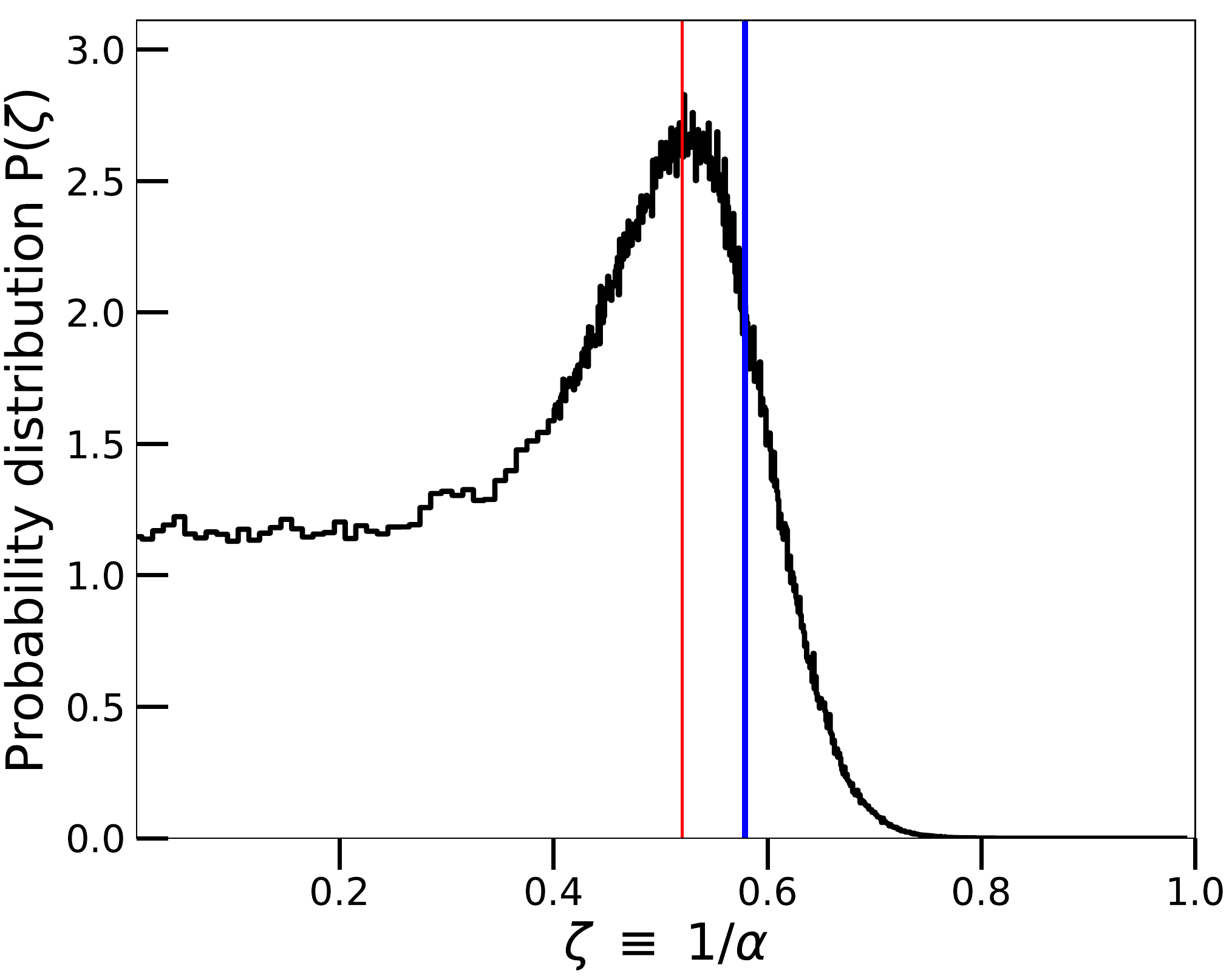}
    \includegraphics[width=0.72\columnwidth]{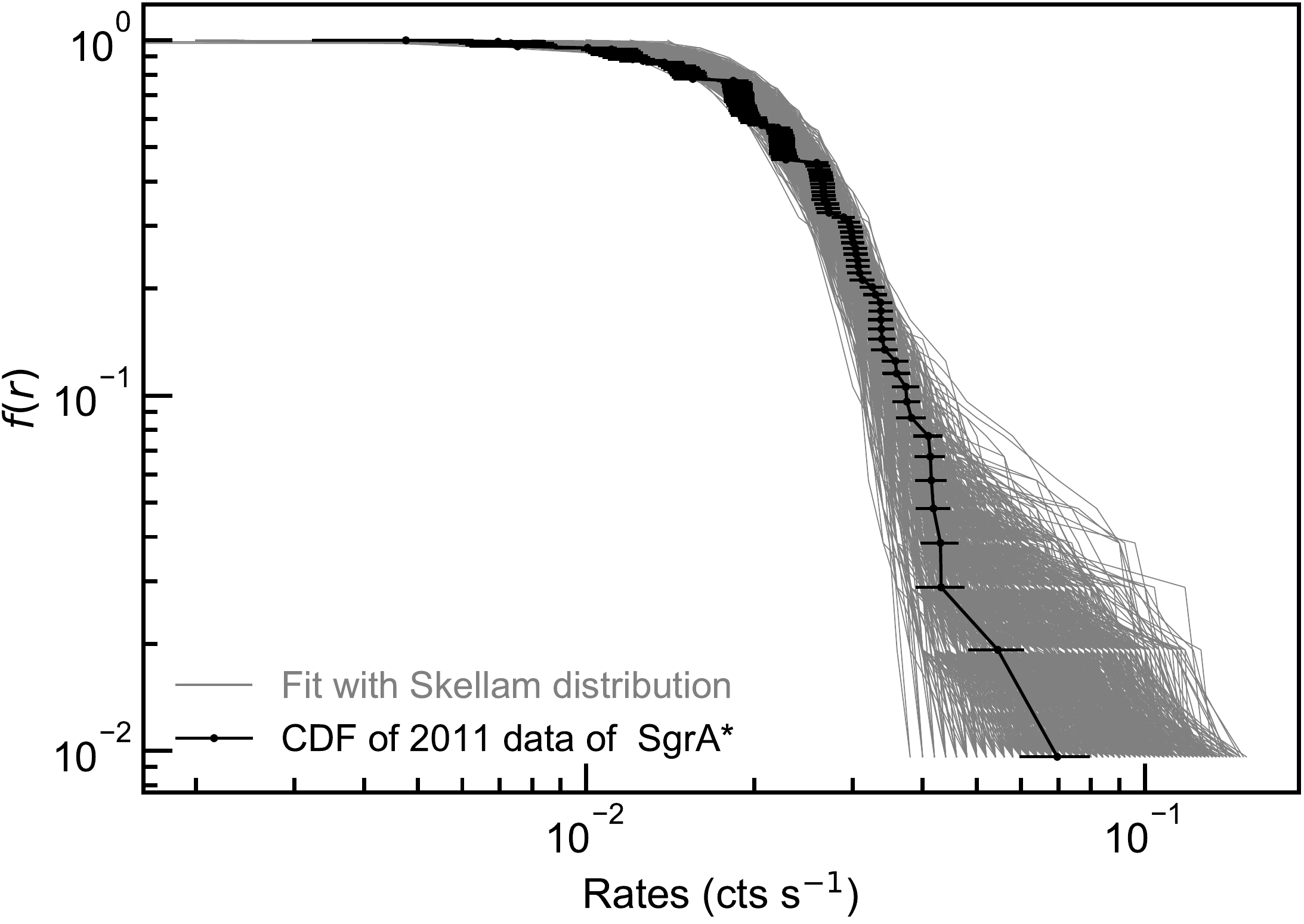}
    
    \includegraphics[width=0.66\columnwidth]{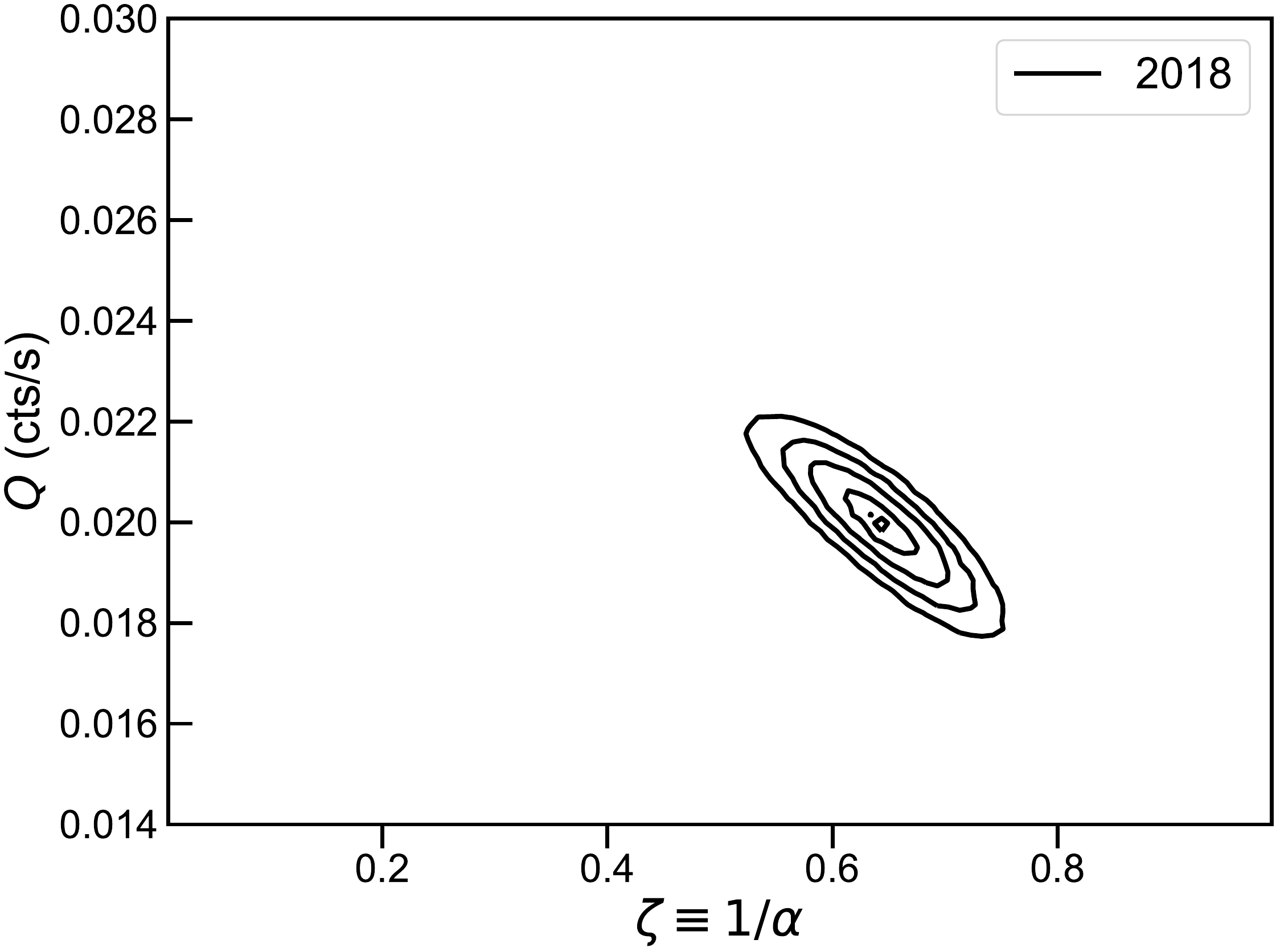}
    \includegraphics[width=0.66\columnwidth]{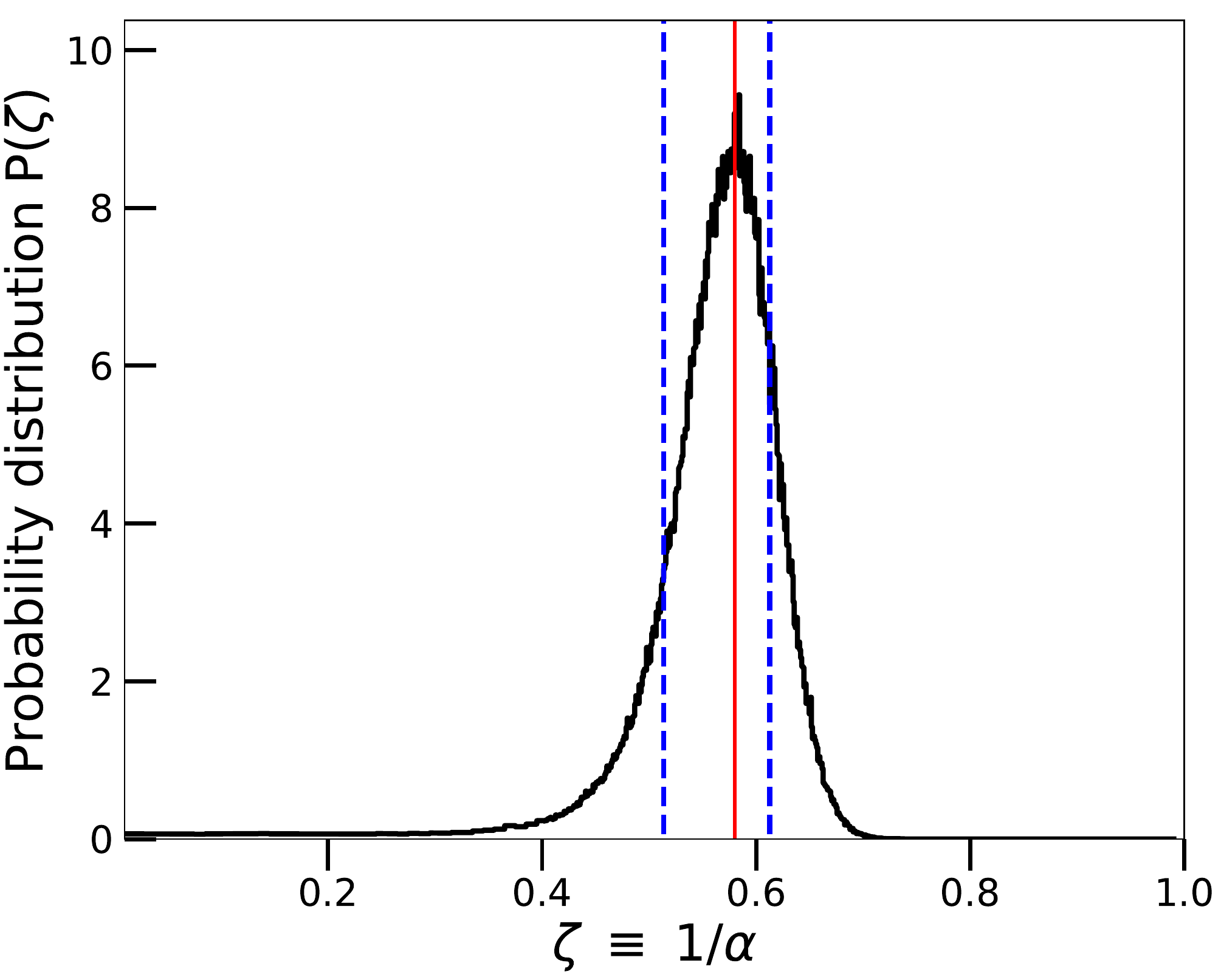}
    \includegraphics[width=0.72\columnwidth]{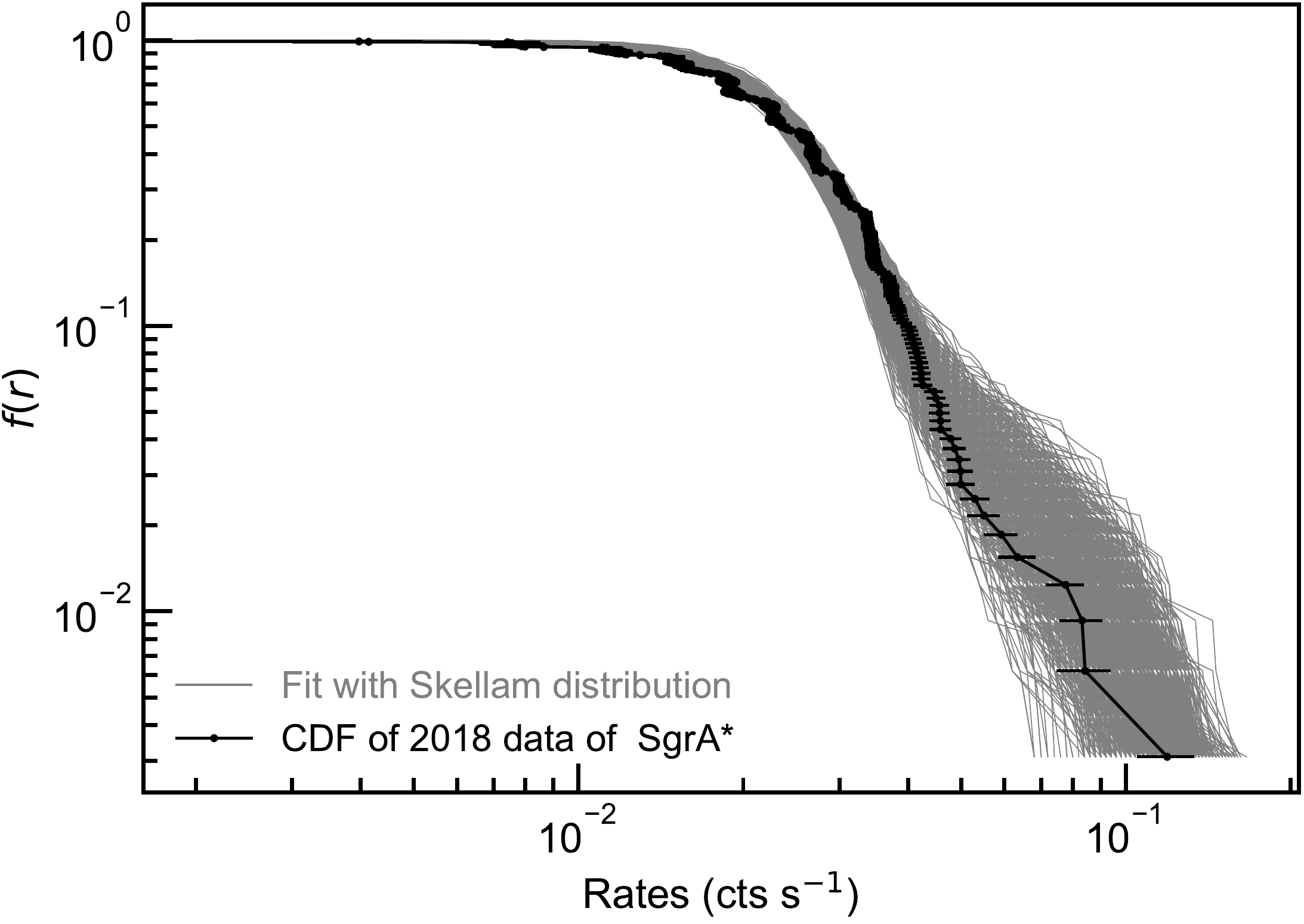}
    
    \includegraphics[width=0.66\columnwidth]{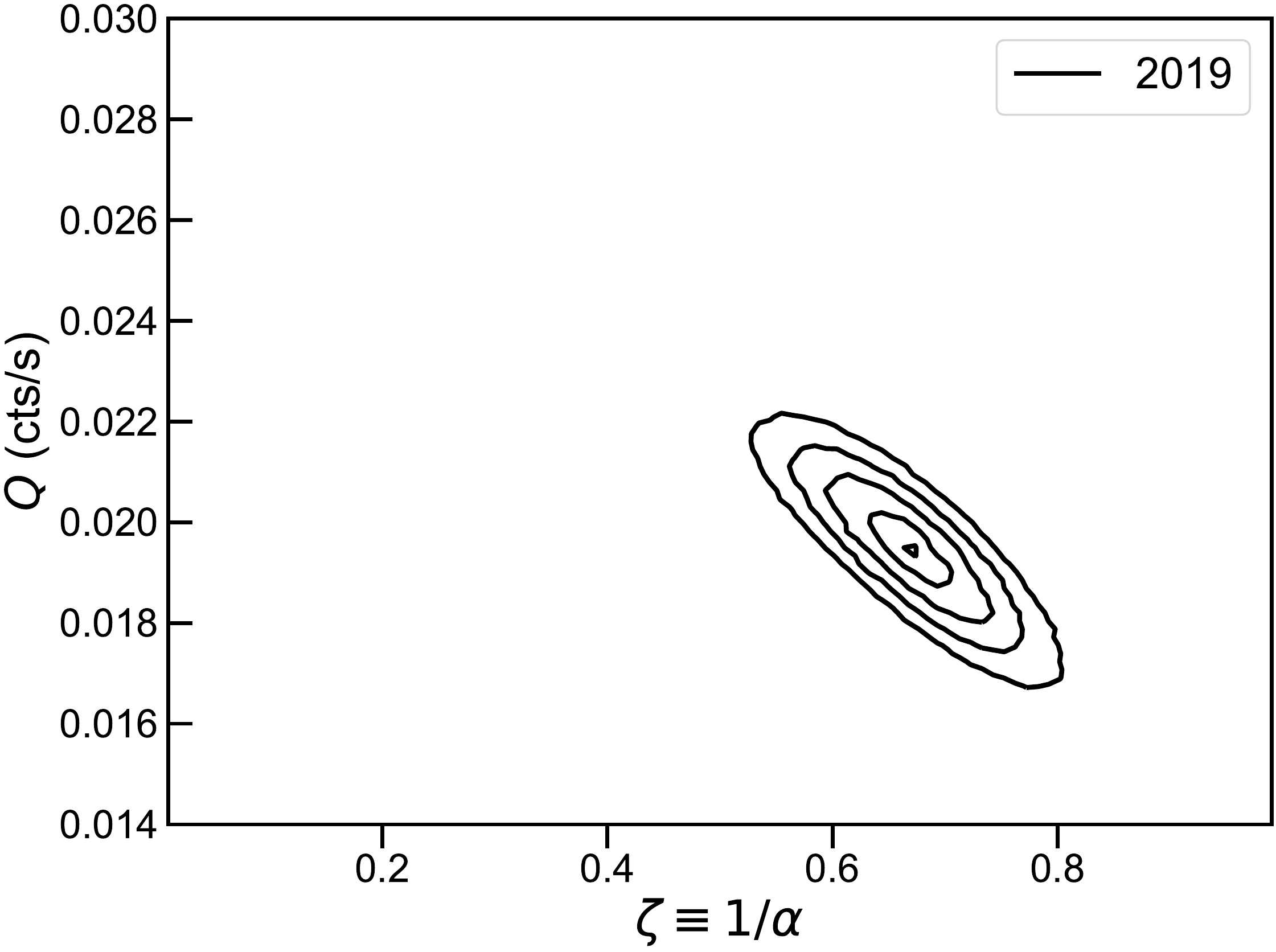}
    \includegraphics[width=0.66\columnwidth]{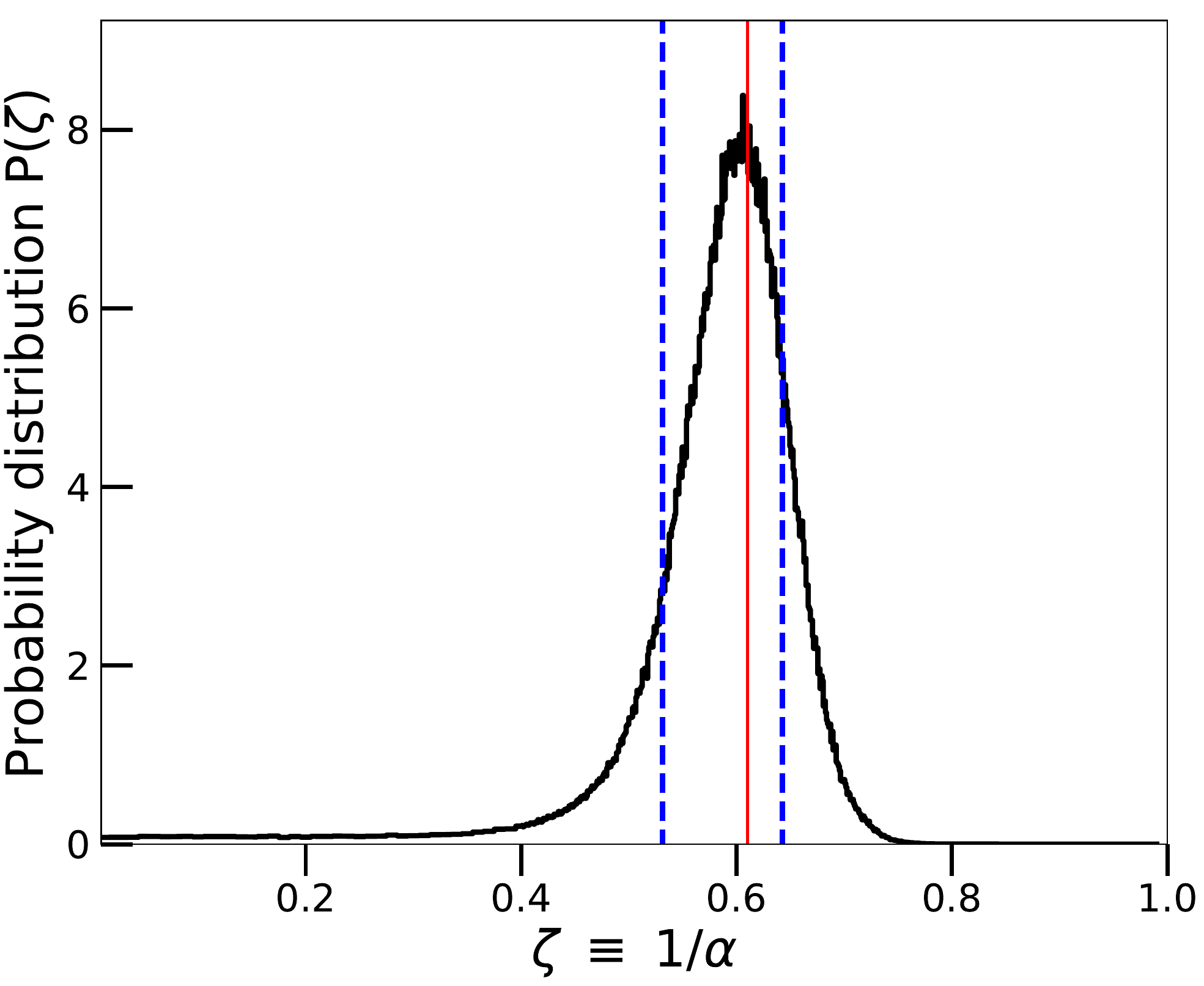}
    \includegraphics[width=0.72\columnwidth]{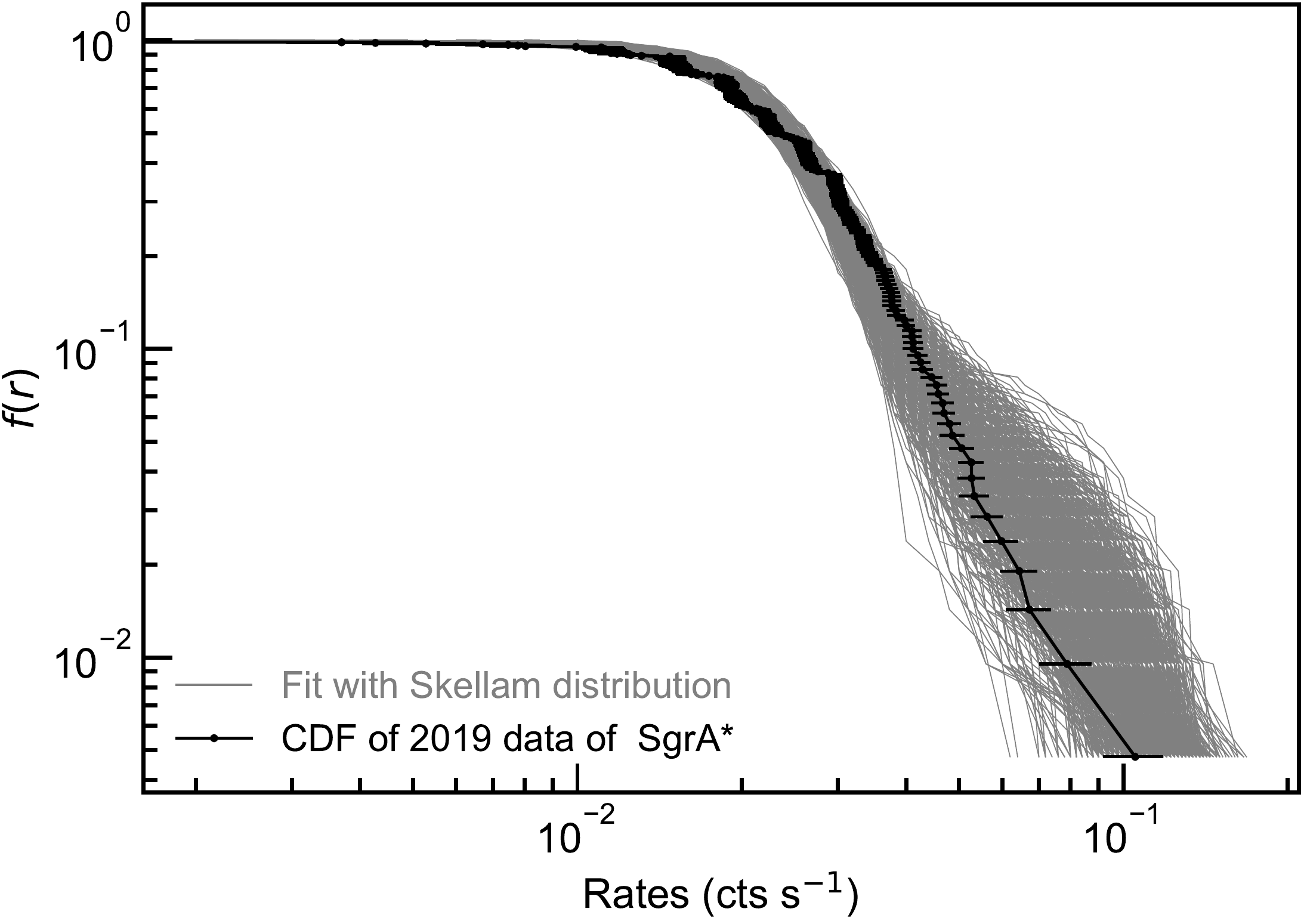}

    \caption{The contour plots in the $\zeta-Q$ plane for the 2D method and probability density function of $\zeta$ for the 1D method, for individual years, and fit of the corresponding CDF. Figures for years 2012 and 2017 are shown in the main body.}
    \label{fig:my_label}
\end{figure*}

\label{lastpage}

\end{document}